\documentclass[journal]{IEEEtran}
\usepackage{comment}
\usepackage{cite}
\usepackage{amsmath,amssymb,amsfonts}
\usepackage{algorithmic}
\usepackage{graphicx}
\usepackage{textcomp}
\usepackage{xcolor}
\usepackage{comment}
\usepackage{multirow}
\usepackage{url}

\ifCLASSINFOpdf
\else

\fi

\begin{document}
\title{Multi-Modal Wireless Flexible Gel-Free Sensors with Edge Deep Learning for Detecting and Alerting Freezing of Gait in Parkinson's Patients}

%
\author{Yuhan~Hou,~\IEEEmembership{Student Member, IEEE}
        Jack Ji, 
        Yi~Zhu,~\IEEEmembership{Member, IEEE} 
        Thomas~Dell, 
        and~Xilin~Liu,~\IEEEmembership{Senior~Member,~IEEE}
\thanks{Y. Hou and X. Liu are with the Department of Electrical and Computer Engineering (ECE), University of Toronto, Toronto, ON, Canada. X. Liu is also with the Toronto Rehabilitation Institute (TRI), University Health Network (UHN), Toronto, ON, Canada. }
\thanks{J. Ji was with the ECE department at the University of Toronto. He is currently with Meta Inc., Seattle, WA, United States.}
\thanks{Y. Zhu was with the ECE department at the University of Toronto and the TRI, UHN. He is currently with Synaptive Medical Inc., Toronto, ON, Canada.}
\thanks{T. Dell is with the Institute of Biomedical Engineering at the University of Toronto.}
}

\markboth{Accepted for publication in the IEEE Transactions on Biomedical Circuits and Systems}
{Shell \MakeLowercase{\textit{et al.}}: Bare Demo of IEEEtran.cls for IEEE Journals}

\maketitle

\begin{abstract}
Freezing of gait (FoG) is a debilitating symptom of Parkinson's disease (PD). This work develops flexible wearable sensors that can detect FoG and alert patients and companions to help prevent falls. FoG is detected on the sensors using a deep learning (DL) model with multi-modal sensory inputs collected from distributed wireless sensors. Two types of wireless sensors are developed, including: (1) a C-shape central node placed around the patient's ears, which collects electroencephalogram (EEG), detects FoG using an on-device DL model, and generates auditory alerts when FoG is detected; (2) a stretchable patch-type sensor attached to the patient's legs, which collects electromyography (EMG) and movement information from accelerometers. The patch-type sensors wirelessly send collected data to the central node through low-power ultra-wideband (UWB) transceivers. All sensors are fabricated on flexible printed circuit boards. Adhesive gel-free acetylene carbon black and polydimethylsiloxane electrodes are fabricated on the flexible substrate to allow conformal wear over the long term. Custom integrated circuits (IC) are developed in 180 nm CMOS technology and used in both types of sensors for signal acquisition, digitization, and wireless communication. A novel lightweight DL model is trained using multi-modal sensory data. The inference of the DL model is performed on a low-power microcontroller in the central node. The DL model achieves a high detection sensitivity of 0.81 and a specificity of 0.88. The developed wearable sensors are ready for clinical experiments and hold great promise in improving the quality of life of patients with PD. The proposed design methodologies can be used in wearable medical devices for the monitoring and treatment of a wide range of neurodegenerative diseases. 
\end{abstract}


\IEEEpeerreviewmaketitle

\section{Introduction}

\begin{figure}[!ht]
    \centering
    \includegraphics[width=1\linewidth]{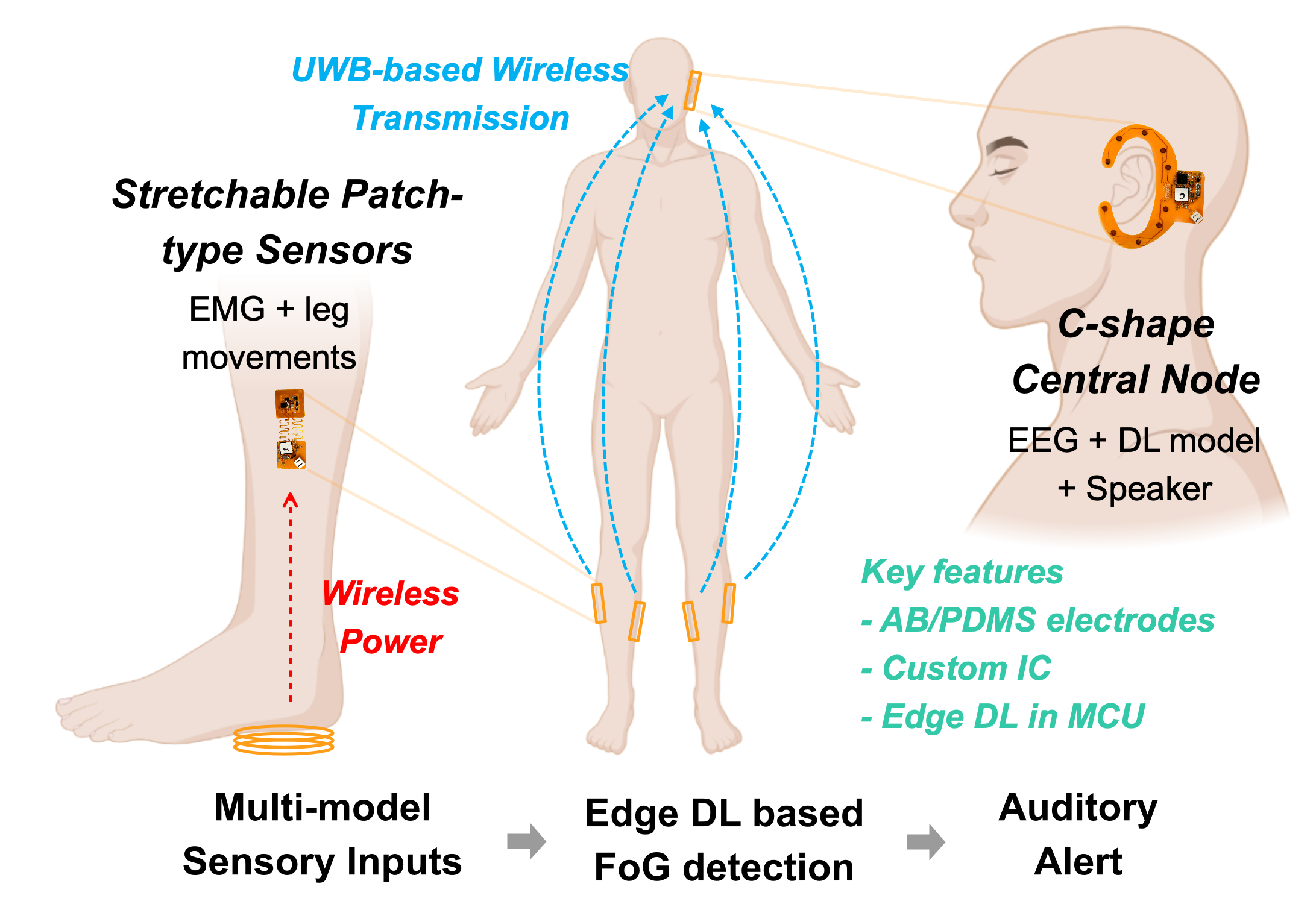}
    \caption{Illustration of the overall FoG detection system consists of wireless gel-free sensor nodes. Four stretchable patch-type sensors are placed on the gastrocnemius and tibialis anterior muscles of the patient's both legs to collect EMG and leg movement, and wirelessly send the data to the C-shape central node. The central node is placed around the patient's ear, which measures EEG and detects FoG using an on-device DL model with multi-modal sensory inputs. An auditory alert will be generated when a FoG event is detected. }
    \label{fig:intro}
\end{figure}

\IEEEPARstart{P}{arkinson's} disease (PD) is a prevalent neurological disorder that affects more than ten million patients worldwide \cite{bloem2021parkinson}. Age is a major risk factor for the development and progression of PD \cite{hindle2010ageing}. With the global aging population, the prevalence of PD will be worse in the coming decades. Currently, there is no cure for PD, but only treatments to reduce symptoms. One of the most common symptoms among PD patients is freezing of gait (FoG), which is defined as sudden, short, and temporary episodes of an inability to move the feet despite the intention of walking \cite{limousin2019long}. More than 79.2\% of PD patients in advanced states suffer from FoG \cite{tan2011freezing}. Detecting FoG in real time and alerting the patient and their companions using wearable devices can help prevent falls, which will significantly improve the life qualities of PD patients.

Detecting and alerting FoG in real time requires wearable sensors to extract FoG-relevant physiological and physical signals, along with algorithms that can predict the onset time of FoG based on collected sensory signals, as illustrated in Fig. \ref{fig:intro}. To allow long-term use outside clinical settings, devices must be comfortable and easy to wear, and the algorithm needs to be executed on the devices. Although cloud computing permits the use of extensive computational resources, it suffers from latency due to data transmission, prevents offline use, and poses cybersecurity concerns during data transmission \cite{liu2021edge}. Recently, machine learning (ML)-based methods, including deep learning (DL) models, have been developed for FoG detection and have shown superior performance over conventional algorithms~\cite{armstrong2020diagnosis,tuauctan2020freezing}. Commonly used sensory signals for FoG detection include electroencephalogram (EEG), electromyography (EMG), and patients' movement patterns, especially leg movement \cite{zhang2022multimodal}. However, existing DL models often have high computational demands, preventing deployment in low-power wearable devices.

In this paper, we report the development of flexible multi-modal wireless sensors for detecting and alerting FoG using edge DL models. Fig. \ref{fig:intro} illustrates the overall system, which consists of two types of wireless sensors: (1) a C-shape central node placed around the ears of the patients to collect the EEG signal and generate an auditory alert when FoG is detected; (2) stretchable patch-type sensors attached to the legs of the patients to collect EMG and leg movement and wirelessly send the data to the central node. 
{\color{black}Compared to conventional cap EEG headset, the proposed C-share around ear EEG recording device is much smaller, lighter, and less cumbersome, making them easier to wear for extended periods without discomfort. Additionally, earlobe is much less sensitive than the scalp, which makes ear EEG devices less intrusive and more comfortable to use for patients, espeically for applications such as the detection and alert of FoG. To the best of our knowledge, FoG detection using ear EEG signal has not been explored, despite evidences suggest that ear EEG signals are strongly correlated to cap EEG signals \cite{mikkelsen2015eeg,meiser2020sensitivity,kidmose2013study}. This work's development of wearable sensors represents a significant advancement towards the validation of the clinical effectiveness of detecting FoG and other neurological disorders in future clinical trials.}

All sensors are fabricated using commercial flexible printed circuit board (F-PCB) technology. For recording EEG and EMG, adhesive gel-free acetylene carbon black (AB) and polydimethylsiloxane (PDMS) electrodes are fabricated directly on the flexible substrate of the sensors to allow conformal contact on the skin for long-term wear. Leg movements are measured using a 3-axis MEMS accelerometer (ACC). Custom integrated circuits (ICs) are designed and used for the acquisition and digitization of sensory signals and wireless data transmission. A chopper-stabilized low-noise amplifier is developed for EEG recording. A 12-bit analog-to-digital converter (ADC) is developed with a hybrid digital-to-analog converter (DAC) for digitizing sensory signals. Ultra-wideband (UWB) transmitter (TX) and receiver (RX) are developed for low-latency wireless data transmission. A DL model is developed for FoG detection using multi-modal sensory input. The trained model is trained offline and deployed into a low-power general-purpose microcontroller (MCU) for real-time FoG prediction. Thanks to the custom IC design, the patch-type sensor achieves ultra-low power consumption and can be wirelessly powered by inductive coupling. This permits a battery-free design that minimizes the weight and thickness of the sensor.

The main contributions of this work are as follows. First, we developed a first-of-its-kind flexible wireless sensor network for FoG detection and closed-loop auditory alert. Second, we developed a novel lightweight edge DL model for FoG detection with multi-modal signal inputs. The model has been successfully deployed in a low-power resource-restricted MCU and achieved a detection performance comparable to the state-of-the-art works. Third, we designed and validated custom ICs and gel-free AB/PDMS electrode fabrication techniques for wearable sensors. The custom IC designs achieve reliable high-precision signal acquisition and low power wireless data transmission; the low-cost electrode fabrication is suitable for conformal wearable sensors. {\color{black}} The design methods and techniques can be used in a wide range of applications.

The rest of the paper is organized as follows. Section II introduces the electrode fabrication method and the design of the flexible sensors. Section III discusses the development of the DL model. Section IV presents the implementation of the custom IC and peripheral circuits. Section V shows the experimental results. Finally, Section VI concludes the paper.

\section{Electrode and Flexible Sensors}

\subsection{AB/PDMS Electrode Fabrication}

\begin{figure}[!ht]
    \centering
    \includegraphics[width=.85\linewidth]{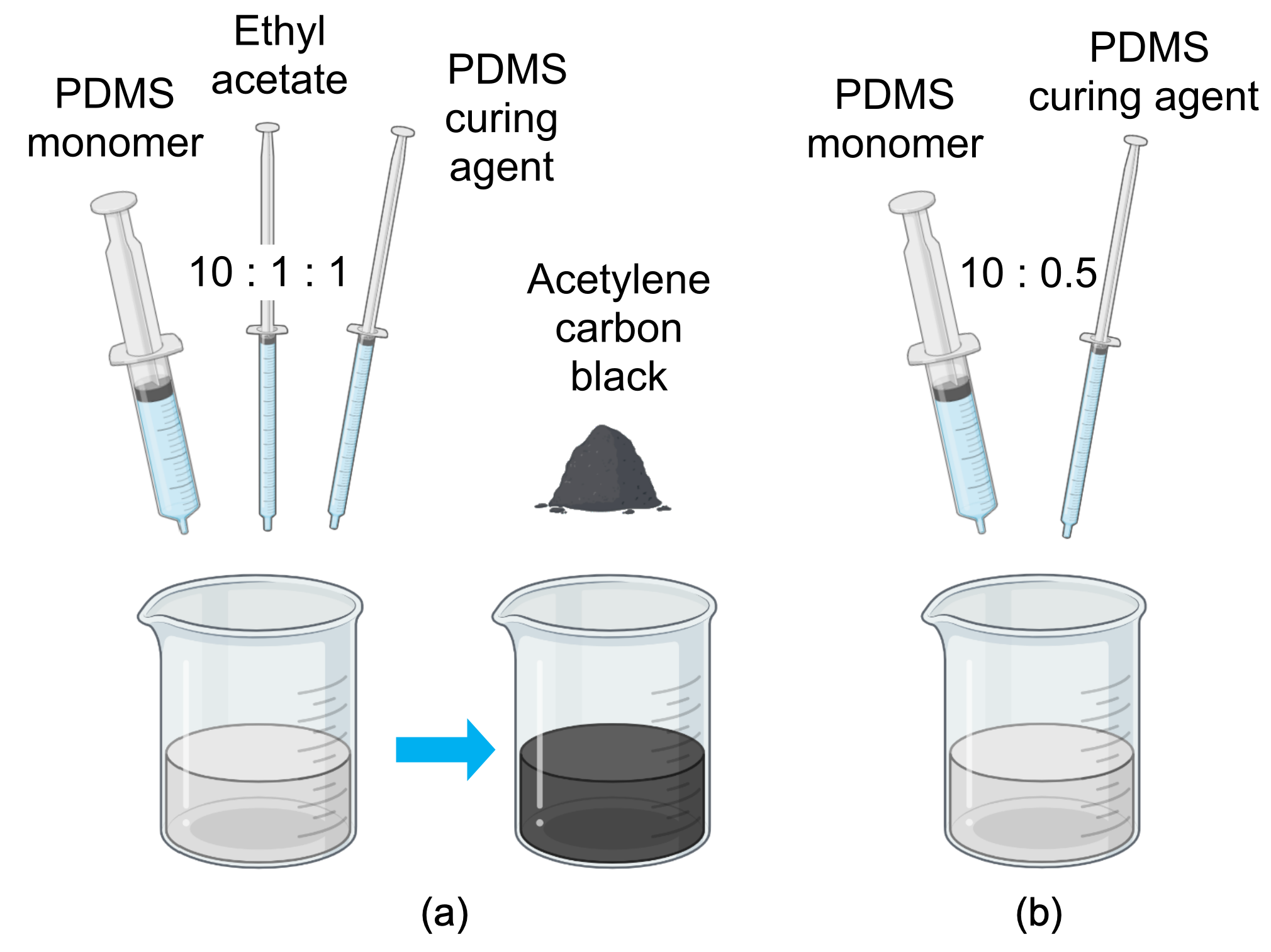}
    \caption{Illustration of the material preparation process of the electrodes. The process (a) is for the AB/PDMS electrode contacts, and the process (b) is for the adhesive non-conducting part of the pads. The processes are conducted at 25 \textdegree C.}
    \label{fig:fab_process}
\end{figure}

Conductive electrolyte gel is commonly used in wearable sensors for reducing electrode impedance and improving signal quality in the measurement of electrophysiological signals, such as EMG, EEG, and ECG. However, electrolyte gel is not comfortable and stable for long-term use and may cause skin irritation \cite{reyes2014novel,jung2012cnt}. In this work, we fabricate AB/PDMS electrodes that are both adhesive and conductive without silicon-based adhesive gel \cite{dong2021fully,reyes2014novel,liu2017wearable}. Fig. \ref{fig:fab_process} illustrates the preparation process of the electrodes.

Electrode contacts are made of PDMS monomer, curing agent (Silicon Elastomer), and ethyl acetate (EtOAc, 99.8\%, Sigma-Aldrich) in a beaker at a ratio of 10:1:1, with AB particles (Carbon black, acetylene, 99.9+\%, 50\% compressed, Thermo Scientific) at 5 wt\%. The rest of the electrode pads are filled with PDMS monomer and curing agent, which is mixed in a beaker at a ratio of 10:0.5. Both mixtures were deposited on the F-PCB substrate and evacuated for 12 hours at 25 \textdegree C before use. The preparation process is simple, and the fabricated AB/PDMS electrodes are comfortable to wear over the long term. The electrode impedance is comparable to conventional gel electrodes. The measurement results are discussed in Section \ref{sec_exp}.

\subsection{C-shape Central Sensor}
\begin{figure}[!ht]
    \centering
    \includegraphics[width=1\linewidth]{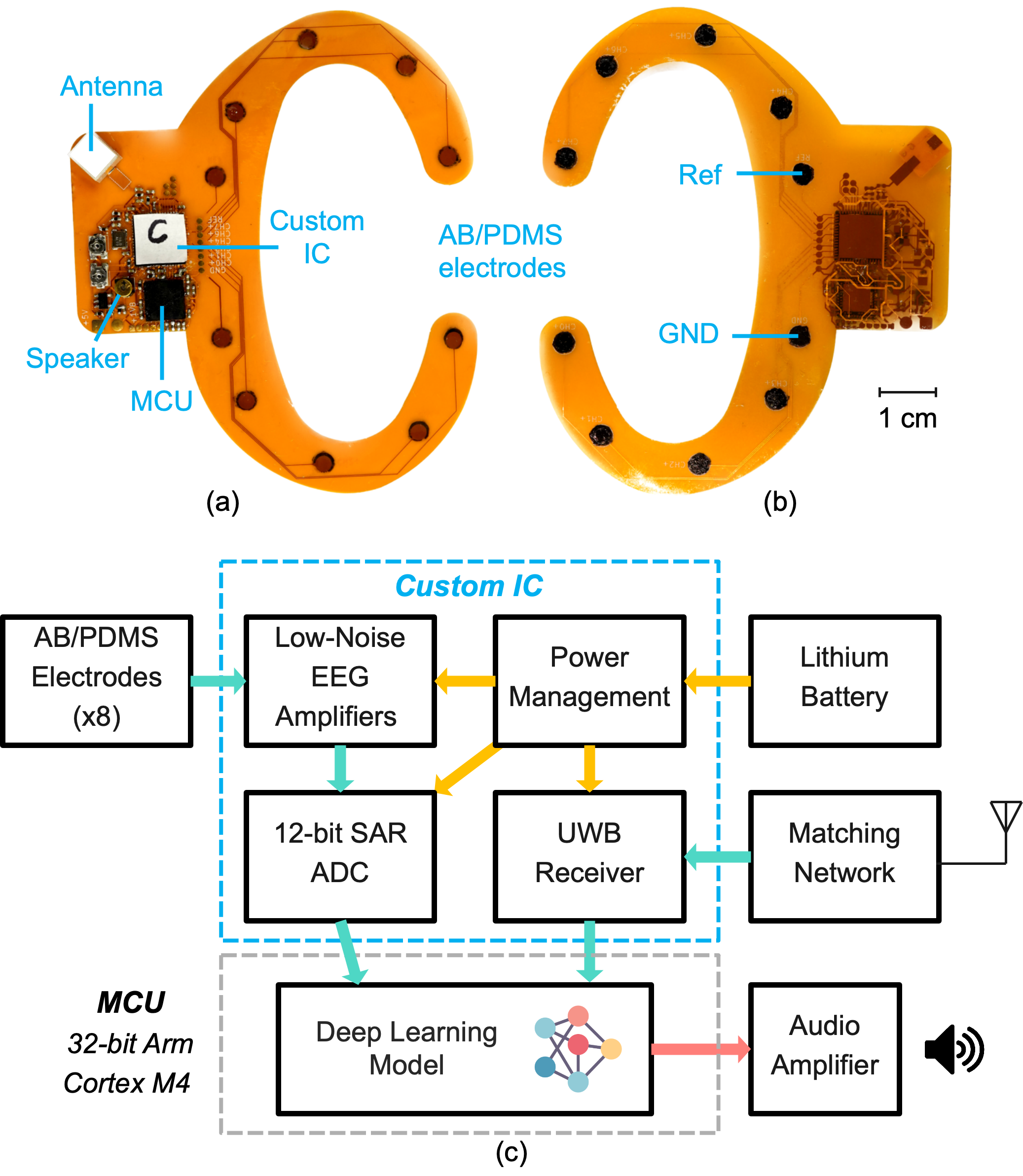}
    \caption{(a) and (b) shows the front and back photos of the assembled C-shape central sensor, respectively. (c) The high-level block diagram of device.}
    \label{fig:Cshape}
\end{figure}

The C-shape central sensor is fabricated on an F-PCB. Figs. \ref{fig:Cshape}(a) and (b) show the front and back of the assembled C-shape central sensor, respectively. The high-level block diagram of the sensor is shown in Fig. \ref{fig:Cshape}(c). The sensor consists of ten AB/PDMS electrodes, a custom IC, a low-power MCU (nRF52840, Nordic Semiconductor), an audio amplifier with a speaker, a rechargeable lithium battery, and peripheral circuits. The electrode placement follows the design of the open-source concealed around-the-ear EEG electrodes (also known as cEEGrids) design \cite{debener2015unobtrusive,knierim2021open}. Eight of the ten electrodes are used to measure EEG signals, one is used as the reference electrode, and one as the ground electrode (Fig. \ref{fig:Cshape}).

The custom IC consists of eight low-noise neural amplifiers, which share one 12-bit successive approximation register (SAR) ADC. An UWB receiver is designed to collect data from the patch-type sensors. All measured and received signals are sent to the MCU for processing. The central sensor is powered by a 3 V lithium battery. A speaker is integrated to generate auditory alerts. An on-board switching regulator is used to convert the battery voltage to 1.8 V for powering the custom IC. Low dropout regulators (LDOs) on-chip are used to generate clean 1.8 V and 1.2V supply voltages to the analog, digital, and radio-frequency (RF) circuits on-chip. The circuit design details of the custom IC are discussed in Section \ref{sec_IC}.

\subsection{Stretchable Patch-type Sensor}

\begin{figure}[!ht]
    \centering
    \includegraphics[width=1\linewidth]{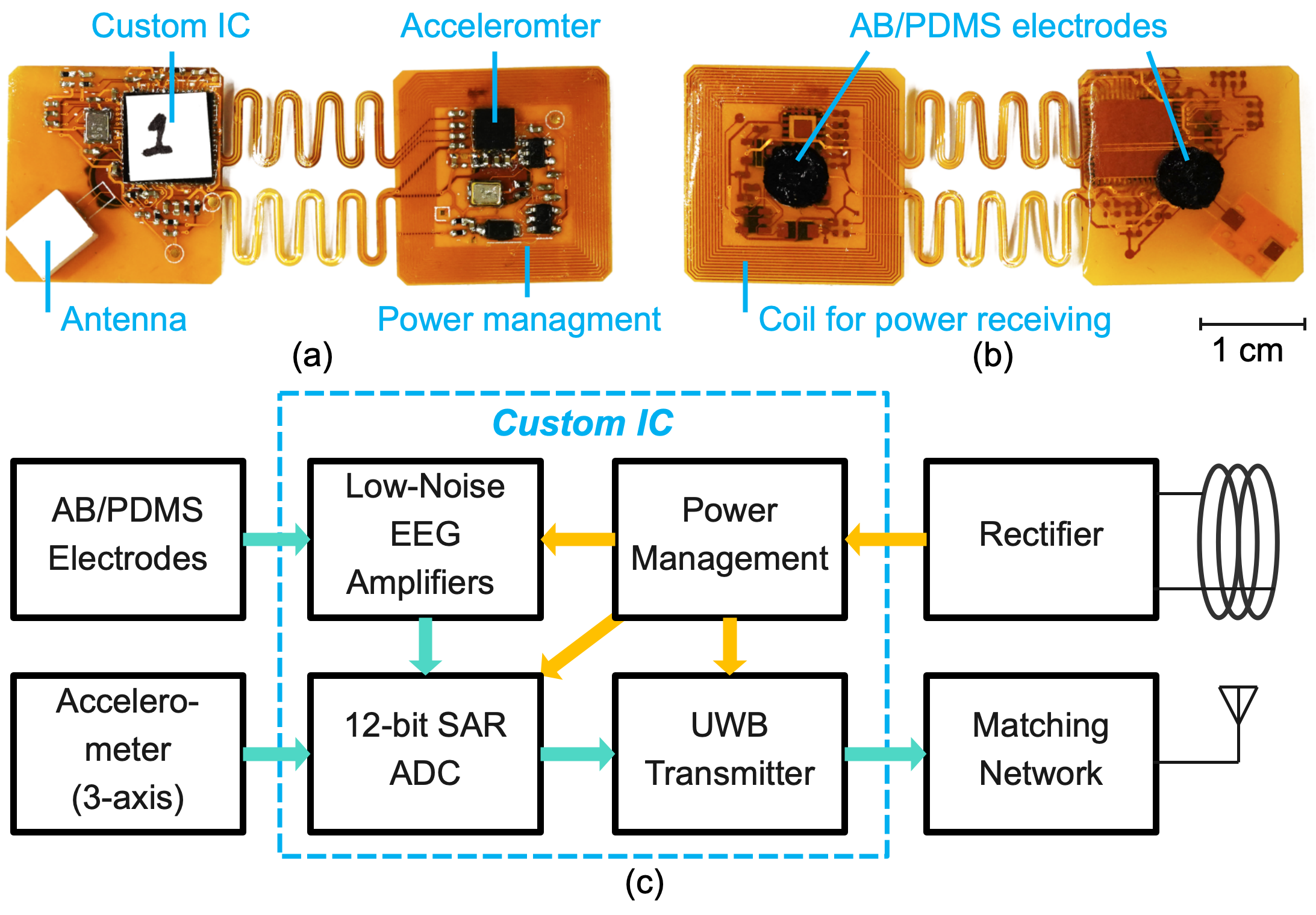}
    \caption{(a) and (b) shows the front and back of the developed stretchable patch-type sensor, respectively. (c) The high-level block diagram of the device.}
    \label{fig:patch_sensor}
\end{figure}

Figs. \ref{fig:patch_sensor}(a) and (b) show the front and back of the developed stretchable patch-type sensor, respectively. The high-level block diagram of the sensor is shown in Fig. \ref{fig:patch_sensor}(c). The sensor consists of two AB/PDMS electrodes, a custom IC, a 3-axis MEMS accelerometer, and an inductive coil with power management circuits. 
Four stretchable patch-type sensors will be placed on the gastrocnemius and tibialis anterior muscles of both legs of the patient \cite{zhang2022multimodal}, as illustrated in Fig. \ref{fig:intro}. The serpentine-shaped structure connects the two patches, which permits reliable attachment during muscle contraction and relaxation. {\color{black}In this work, a monopolar configuration is employed for EMG recording, in which the reference and ground signals are acquired using the same electrode. Alternatively, a bipolar recording can be performed by adding an extra electrode to serve as a dedicated ground.}

One channel of the low-noise amplifier in the custom IC is used for recording EMG signal, which is digitized by the on-chip ADC. The accelerometer (ADXL356, Analog Devices Inc.) is used to detect movements of the legs. The ADXL356 has three analog outputs for the ACC in the three axes with a full-scale range of $\pm$10 g. The analog outputs are also digitized by the on-chip ADC. All digitized data is sent to the central sensor wirelessly via an on-chip UWB transmitter. 

To maximize the comfort of wear and long-term use, the weight and thickness of the patch-type sensors need to be minimized. In this work, we adopt inductive wireless powering to permit battery-free operation. The wireless powering is through inductive coupling. A rectifier circuit is integrated to recover DC voltage, followed by a switching regulator for powering the custom IC at 1.8 V.

\section{DL Model Development}

The DL model was developed and trained using a public FoG dataset \cite{zhang2022multimodal}. {\color{black}This study was conducted in Beijing Xuanwu Hospital, China, and obtained ethical approval (No. 2019-014) from the Ethics Commitee of Xuanwu Hopsital, Capital Medical University, Beijing, China. The raw data of the database is publicly available through this link \cite{data2021}.} The dataset contains multi-modal sensory data collected from 12 PD patients. {\color{black}All patients were asked to do four walking tasks during the study. To obtain valid data containing sufficient FoG episodes, data were collected in the off-medication state of patients. All patient identifiable information has been removed from the database before sharing. Our work relies exclusively on the anonymous information and is exempt from the ethical review based on the guidance of the human research ethics board (REB) of the University of Toronto.} Four EEG channels were used as input for the model instead of all 25 channels that are available in the dataset. 
3-axis ACC signals and EMG signals from two sensors placed on the tibialis anterior muscles are used. The selection of the input signal is a trade-off between the performance, inference latency, and the resources available in the selected MCU.

Conventional ML algorithms using hand-crafted feature selection \cite{zhang2022multimodal} and DL models \cite{lin2022edge,guo2022high,lin2023wireless} have been developed for FoG detection. DL approaches have advantages in extracting hidden features that are often neglected during manual feature engineering and can achieve superior performance over conventional algorithms. In addition, tailoring the model and parameters for each patient's specific situation permits optimal performance given the resources. DL uses an end-to-end optimization framework without human intervention, which reduces the cost of patient-specific training. Therefore, we focused on developing DL models in this work. 

\begin{figure}[!ht]
    \centering
    \includegraphics[width=.9\linewidth]{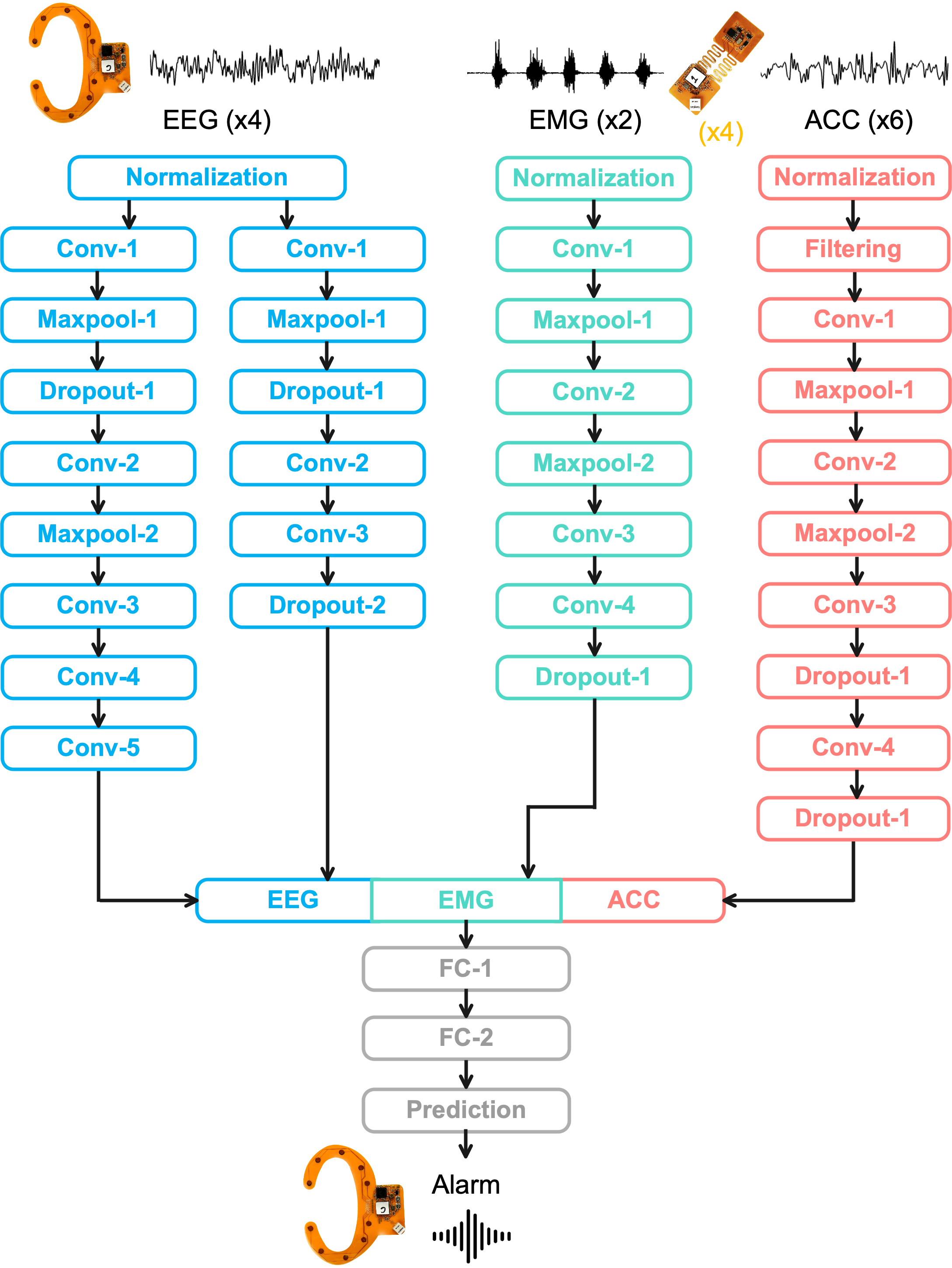}
    \caption{The architecture of the developed multi-modal DL model for FoG detection. Depth-wise CNNs are used for each signal domain. The learned representations are then concatenated for the fully connected layers for the final FoG prediction.}
    \label{fig:dl_model}
\end{figure}

The developed DL model architecture is shown in Fig. \ref{fig:dl_model}. Instead of performing computationally intensive short-time Fourier transform or other time-frequency domain transformation, the model takes time-domain signals directly as input. The input time window is set to 3 s. The first part of the model consists of several depth-wise convolutional neural networks (CNNs) in parallel using time-domain signals as the input directly. CNN is chosen given its strong capability in extracting time-invariant information from input. Two CNN paths with different filter sizes are used for EEG to capture both low-frequency content and high-frequency local signal characteristics \cite{sun2022closed,yao2023cnn}. A single CNN path is used for EMG and ACC, respectively. Each CNN layer is followed by a rectified linear unit (ReLU) function \cite{liu2021edge}. ReLU can effectively resolve the gradient vanishing issue using conventional nonlinear activation functions in deep neural networks \cite{goodfellow2016deep}. ReLU is also energy efficient in computing given its simple implementation. In the second part of the model, the outputs from each domain are concatenated and fed into three fully connected layers for the final FoG prediction.

The DL was developed and trained using Tensorflow \cite{zaccone2017deep}. The training was performed on a Nvidia RTX 4090 GPU with 24 GB GDDR6X memory. The trained model was compressed using quantization and pruning. Quantization reduces the precision of the model parameters for better computational efficiency in limited memory. In this work, we quantize the model from 32-bit floating point numbers to 8-bit integer numbers. Pruning removes parameters of a model that are not essential for the final prediction. Using both quantization and pruning, the model size was reduced from over 2 MB to less than 830 kB, which fits into the memory of the selected MCU model. The performance of the DL model is discussed in Section \ref{sec_exp}.

\section{Circuits Implementation}\label{sec_IC}

\subsection{Low-Noise Chopping Amplifiers and SAR ADC}

Low-noise amplifiers are designed to measure EEG and EMG signals. The low-noise amplifier in this work adopts an architecture with capacitor-coupled inputs to reject DC offsets at the electrode interface. The closed-loop gain of the first-stage of the low-noise amplifier is designed to be 25, which is set by the ratio of the input and feedback capacitors \cite{liu2016design}. A second stage provides an additional programmable gain of 2, 4, 8, or 16. The default total gain setting for EEG and EMG is 200 and 50, respectively. This is designed to best fit the signal within the dynamic range of the amplifier and the ADC \cite{zhang2020electronic}.

\begin{figure}[!ht]
    \centering
    \includegraphics[width=.9\linewidth]{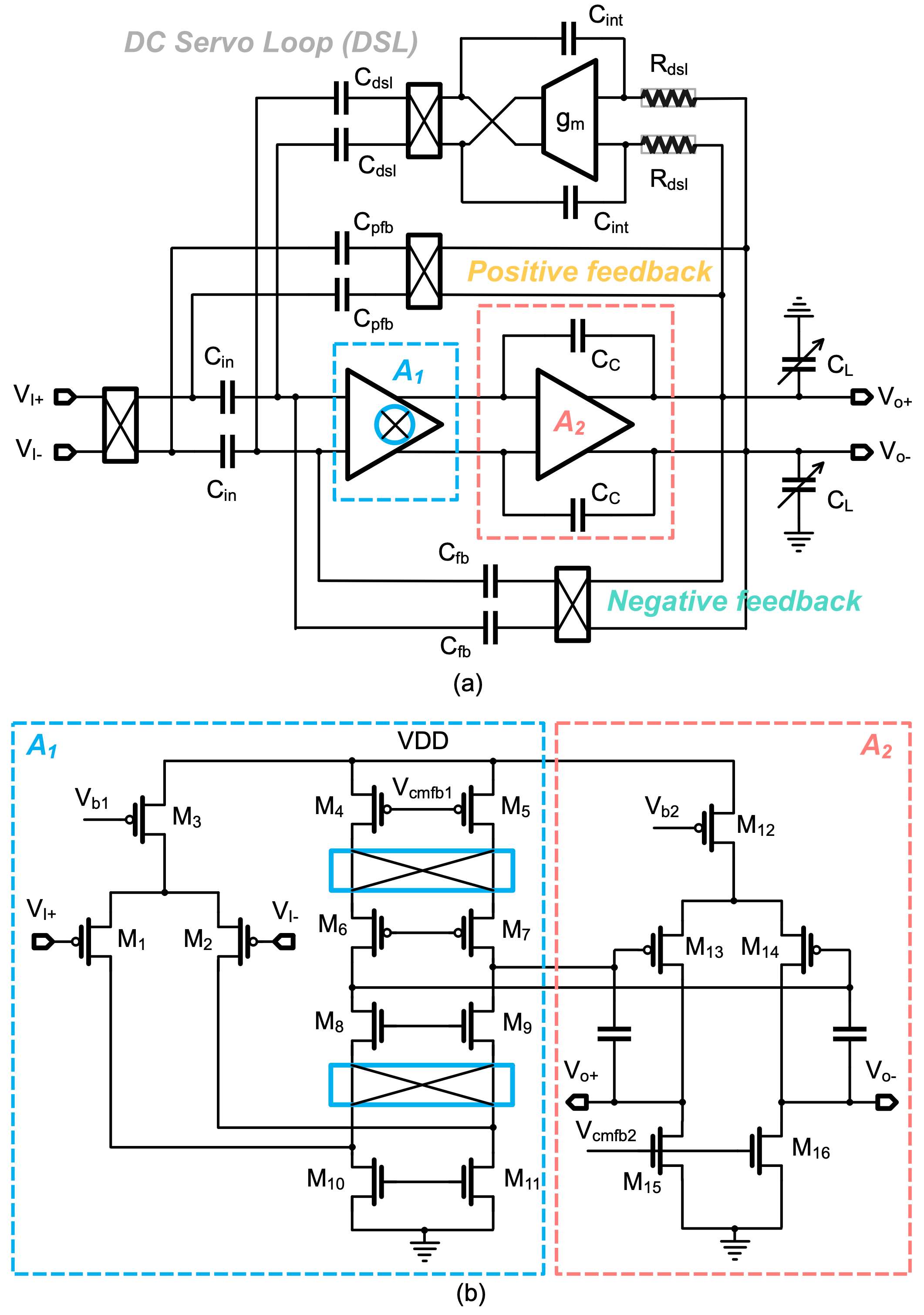}
    \caption{(a) The simplified circuit schematic of the chopper-stabilized neural amplifier with multiple loops. (b) The circuit schematic of the core amplifiers $A1$ and $A2$. }
    \label{fig:ckt_amp}
\end{figure}

The core amplifier uses a fully differential folded-cascode architecture. A bandwidth of 0.5 Hz to 60 Hz is used for EEG recording. Within this frequency range, the flicker noise (i.e., 1/f noise) of transistors is the dominant source of noise. Chopping is an established circuit technique for removing flicker noise by modulating and amplifying the input signals to high frequencies where flicker noise is negligible \cite{denison20072}. However, chopping reduces the input impedance of the amplifiers due to the switching operation. To mitigate this issue, an input impedance boosting technique is used, which recovers and enhances the input impedance by using positive feedback \cite{liu2020fully,reich2021chopped}. When chopping is enabled, a DC servo loop (DSL) is used to set the high-pass frequency corner. The output capacitor $C_L$ can be programmed to change the low-pass frequency corner. Additional gm-C based analog filtering is implemented in the second stage (not shown in Fig. \ref{fig:ckt_amp}) \cite{liu201512}.

\begin{figure}[!ht]
    \centering
    \includegraphics[width=1\linewidth]{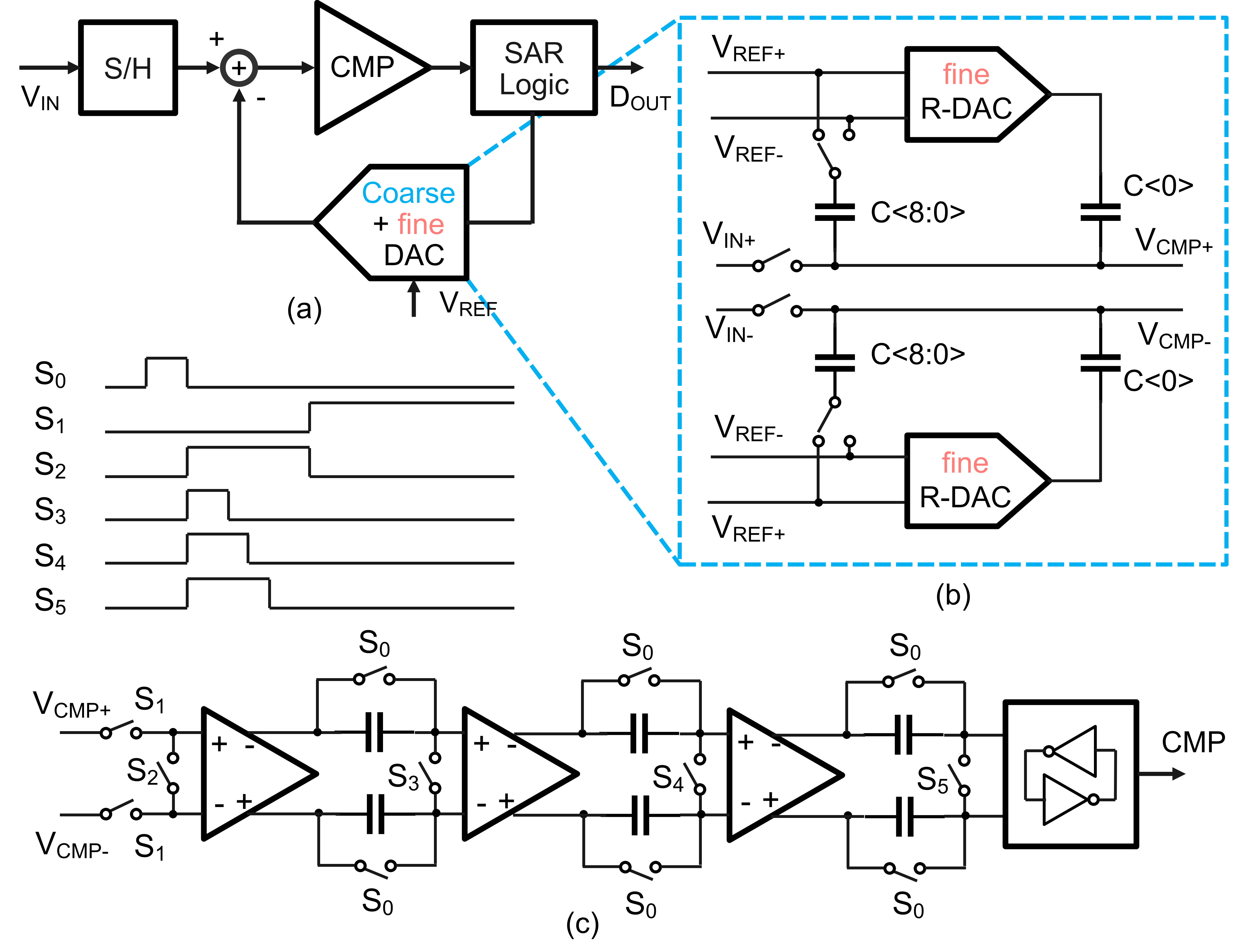}
    \caption{(a) The block diagram of the 12-bit SAR ADC. A single-ended version is used for illustration. The actual implementation is differential. (b) Simplified circuit schematic of the DAC which consists of a 9-bit charge redistribution C-DAC and a 3-bit resistor-string R-DAC. (c) Schematic of the comparator with a 3-stage auto-zeroing pre-amplifier and the timing.}
    \label{fig:ckt_adc}
\end{figure}

A 12-bit SAR ADC is designed to digitize EEG, EMG, and ACC signals. The SAR ADC architecture is chosen given its high power efficiency for moderate to high-resolution analog-to-digital conversion \cite{liu2016design}. A simplified block diagram of the SAR ADC is shown in Fig. \ref{fig:ckt_adc}(a). The key building blocks include a sample and hold (S/H) unit, a hybrid DAC, a comparator, and a SAR logic generation module. The sampling rate of the ADC is designed to be 100~kSps. Although the signal bandwidth and the channel-count in this application are low, a high sampling rate allows the system to go into sleep mode for power saving. The S/H block consists of a bootstrapped switch for high linearity, and the sampling time is 2~$\mu$s. Differential reference voltages at $1/3$ and $2/3$ of the analog supply voltage of 1.8~V are used. The input range of the ADC is thus $\pm$600 mV, resulting in a resolution of $\pm$0.15 mV with an ideal 12-bit ADC. For EEG measurement with a gain of 200, the input range is $\pm$3~mV and the input-referred resolution is $\pm$0.75~$\mu$V; for EMG measurement with a gain of 50, the input range is $\pm$12~mV and the input reference resolution is $\pm$3~$\mu$V.

Capacitive DACs (C-DACs) with charge redistribution are commonly used in low-power SAR ADCs given their superior energy efficiency. However, achieving 12-bit resolution using C-DAC requires a large capacitor array to meet the matching requirement. The large total capacitance often results in high power consumption for the S/H buffer. In this work, a hybrid 12-bit DAC is developed for this SAR ADC, which consists of a 9-bit C-DAC for course conversion and a 3-bit resistive DAC (R-DAC) for the fine conversion. The schematic of the DAC is shown in Fig. \ref{fig:ckt_adc}(b). The minimum size of the unit element is determined by the matching requirement. No calibration is used. The comparator of this ADC is designed to resolve $\pm$0.15 mV. To achieve this precision, we adopt a comparator architecture that consists of a three-stage pre-amplifier and a dynamic latch circuit. The dynamic latch circuit can achieve high gain because of the positive-feedback architecture, but it suffers from mismatches. The three-stage pre-amplifier is designed to provide sufficient gain to compensate for the mismatches of the latch. However, the precision of the pre-amplifier itself is also limited by mismatches. In this work, auto-zeroing is used to reduce the mismatches of the pre-amplifier. Auto-zeroing is a technique that can reduce comparator mismatches by storing and canceling the offsets \cite{verma2007ultra}. The circuit and timing of the comparator and the auto-zeroing operation are illustrated in Fig. \ref{fig:ckt_adc}(c).

\subsection{Ultra Wideband Transceiver}

\begin{figure}[!ht]
    \centering
    \includegraphics[width=.9\linewidth]{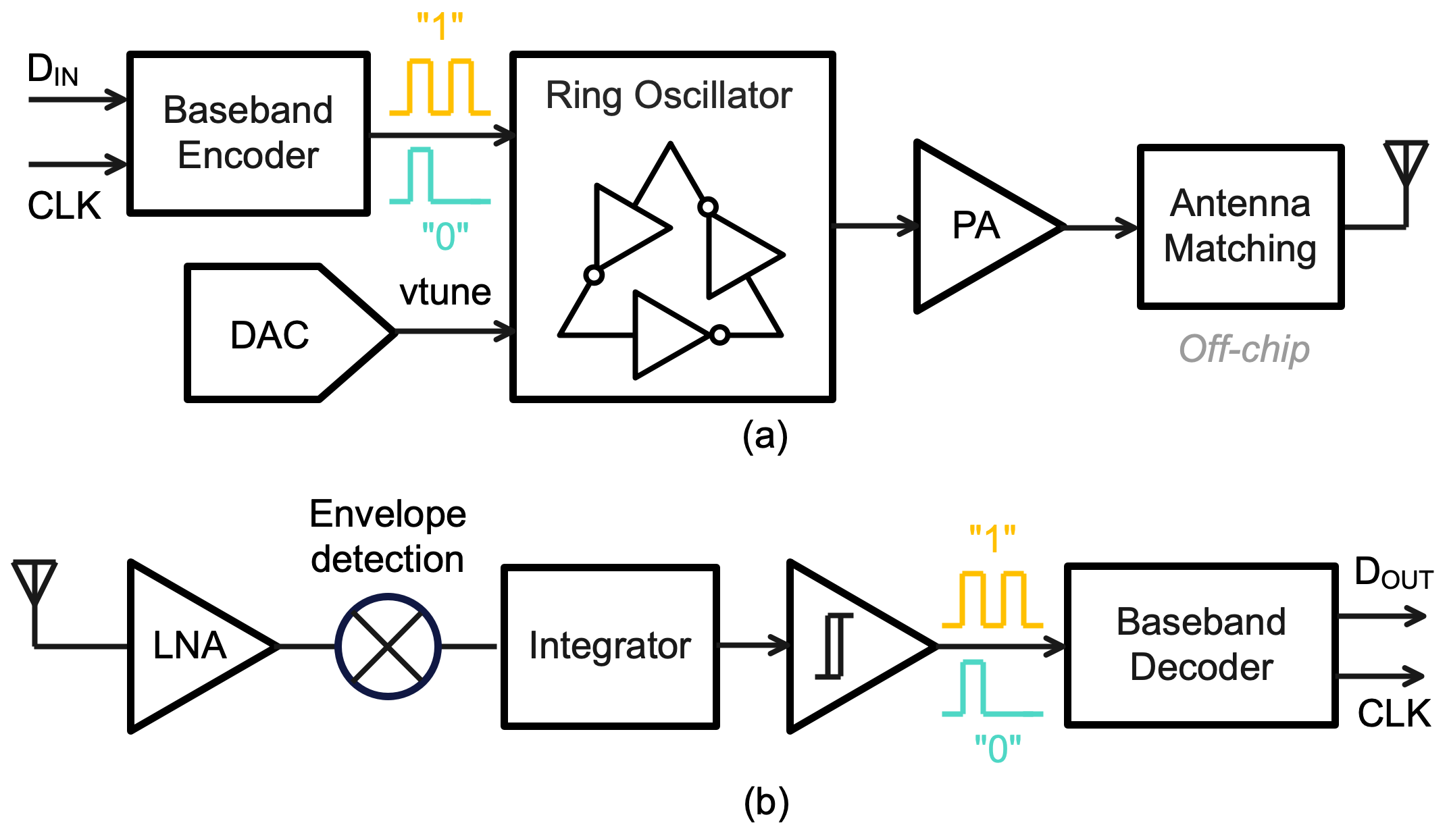}
    \caption{Block diagrams of (a) the UWB TX and (b) the UWB RX.}
    \label{fig:ckt_uwb}
\end{figure}

An impulse radio (IR) UWB transceiver is developed with ON-OFF keying (OOK) modulation. The UWB transceiver is designed to operate from 3.1 to 5 GHz.
A simplified schematic of the UWB TX is shown in Fig. \ref{fig:ckt_uwb}(a). The main building blocks include a baseband encoder, a voltage-controlled ring oscillator (RO), a DAC to generate the control voltage for tuning the frequency of the RO, and a power amplifier (PA). The RO integrates a RC highpass filter and works as the UWB pulse generator \cite{tang2009low}. The OOK modulation is implemented by turning on and off the RO based on the baseband digital bit stream, as illustrated in Fig. \ref{fig:ckt_uwb}. The design achieves low power consumption in a compact layout. 

A simplified schematic of the UWB receiver (RX) is shown in Fig. \ref{fig:ckt_uwb}(b). The key building blocks include a low-noise amplifier (LNA), a RF envelope detector, an energy integrator, a hysteresis comparator, and a baseband decoder. Channelized receiver topology is not used to reduce the design complexity \cite{barras2009low}. The LNA design employs a common-gate architecture with dual feedback to achieve wideband and low noise \cite{kim2010wideband}. The recovered baseband bit stream and clock signal are decoded by an on-chip decoder.

\subsection{Peripheral Circuits}

The reference current of the custom IC is set by an off-chip resistor. The reference is calibrated at the body temperature since it will be worn on the user's body. 
A 16 MHz active oscillator is used to generate clocks for the system. The clocks for different blocks are generated on-chip using frequency dividers and phase-locked loop (PLL) circuits. Communication between the custom IC and the MCU is through a simplified $I^2C$ chip-to-chip communication protocol. The data is buffered in the MCU and then fed into the DL model for inference.  

\begin{figure}[!ht]
    \centering
    \includegraphics[width=1\linewidth]{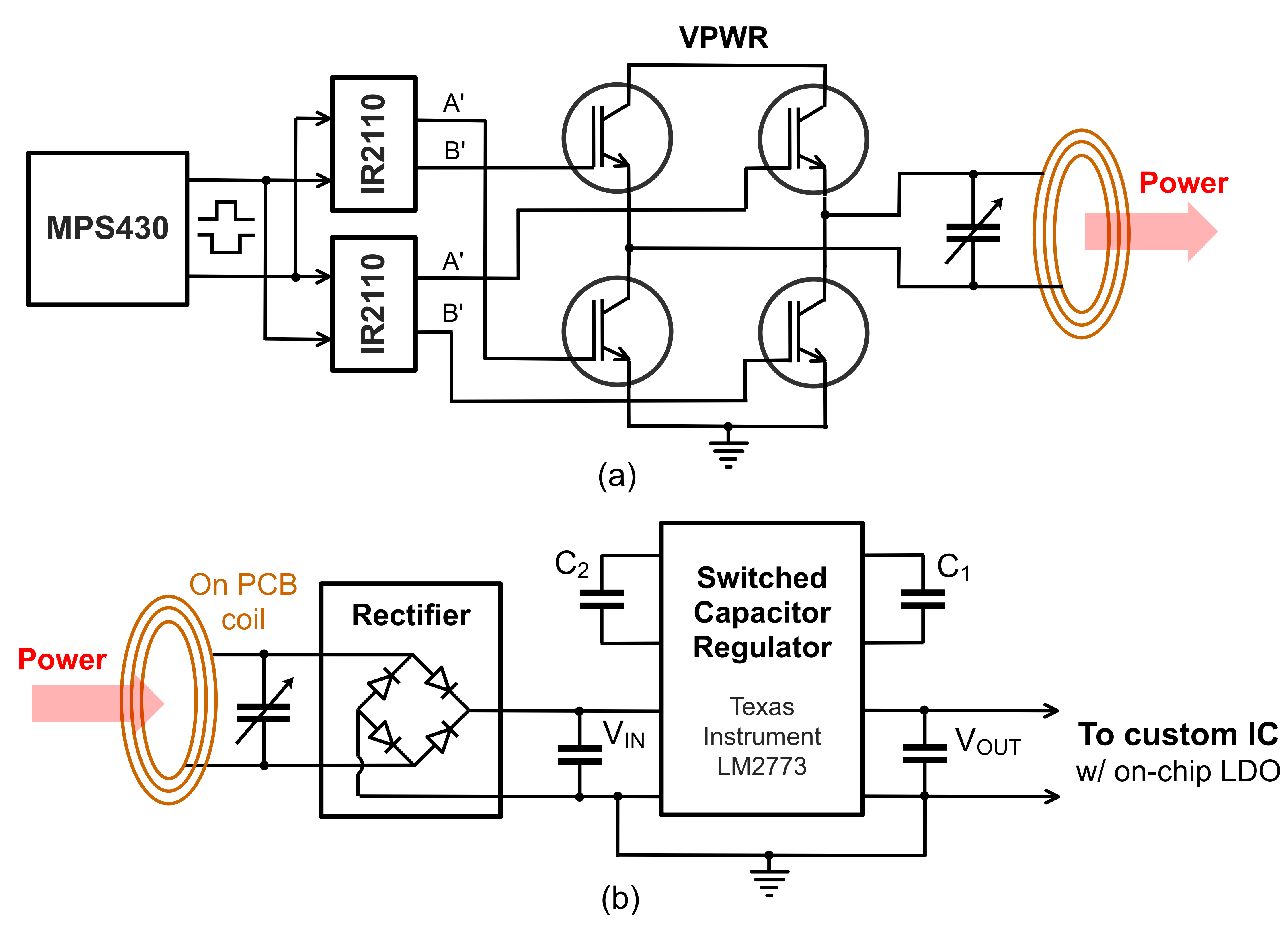}
    \caption{{\color{black}The simplified circuit diagram of (a) the wireless power transmitter, and (b) the wireless power receiver.}}
    \label{fig:wpt}
\end{figure}

Magnetic inductive coupling is used for wireless powering the patch-type sensor. {\color{black}The wireless power transmitter circuit is illustrated in in Fig.~\ref{fig:wpt}(a). The transmitter consists of a battery as the power source, insulated-gate bipolar transistor (IGBT) switches and drivers, an LC tank, and an ultra low-power microcontroller (MSP430, Texas Instruments) for programmable frequency generation. A carrier frequency of 150 kHz is used. Non-overlapping clocks are used to avoid short circuits. A floating channel IGBT driver module (IR2110, International Rectifier) is used to drive the IGBT devices with bootstrap operation \cite{razavi2015bootstrapped}. The wireless power receiver circuit is shown in Fig.~\ref{fig:wpt}(b). The inductive coupling coil is implemented on the F-PCB. To convert the AC resonant waveform to a DC output voltage, a full bridge rectifier comprising four diodes is utilized. A switched capacitor regulator (LM2773, Texas Instruments) is used to regulate the rectified output voltage to 1.8 V. The selected regulator adopted an inductor-less design topology, resulting in a small PCB footprint. The regulated output is sent to the custom IC, where additional LDOs are used to provide low-noise supplies to the analog circuits. We opted to not implement the rectifier on-chip because the AC resonant voltage may well exceed the breakdown voltage of the CMOS process we used. The switched capacitor regulator uses large capacitors (e.g., C1 and C2 in Fig.~\ref{fig:wpt} are 1~$\mu$F) for low output ripple, which are also not suitable for on-chip integration.} 
The user's body serves as the coupling medium and improves efficiency \cite{hao2022wireless}. Accurate modeling of the body channel for power transfer is beyond the scope of this work. 

\section{Experimental Results}\label{sec_exp}

\begin{figure}[!ht]
    \centering
    \includegraphics[width=.9\linewidth]{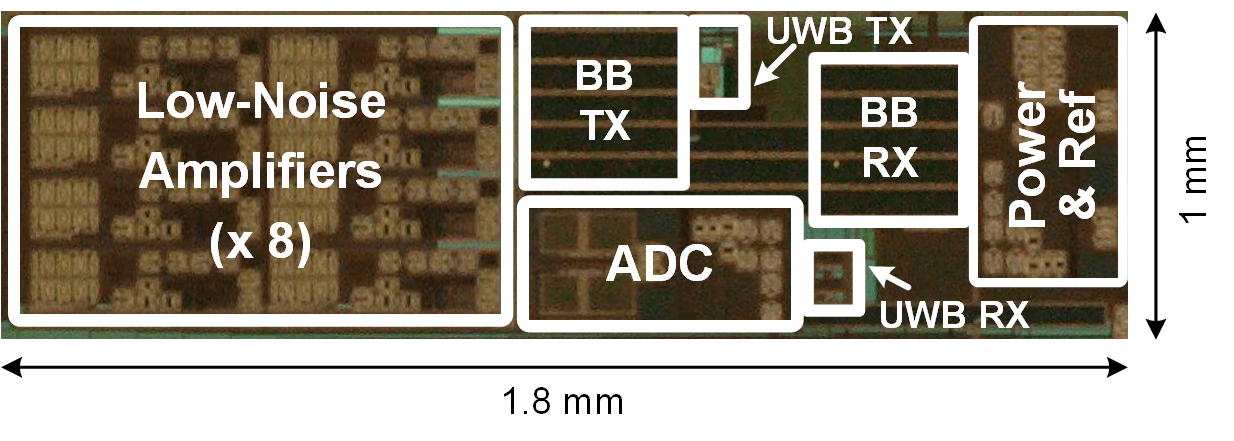}
    \caption{Micrograph of the fabricated custom IC with key blocks highlighted.}
    \label{fig:die_photo}
\end{figure}

A micrograph of a fabricated custom IC in 180 nm CMOS technology is shown in Fig. \ref{fig:die_photo} with the key building blocks highlighted. The design occupies a silicon area of 1.8 mm by 1 mm, excluding the IO pads. The same IC design is used for both the central sensor and the patch-type sensors with different blocks activated. The F-PCB uses Polyimide material with two circuit layers and a thickness of 0.2 mm.

\begin{figure}[!ht]
    \centering
    \includegraphics[width=1\linewidth]{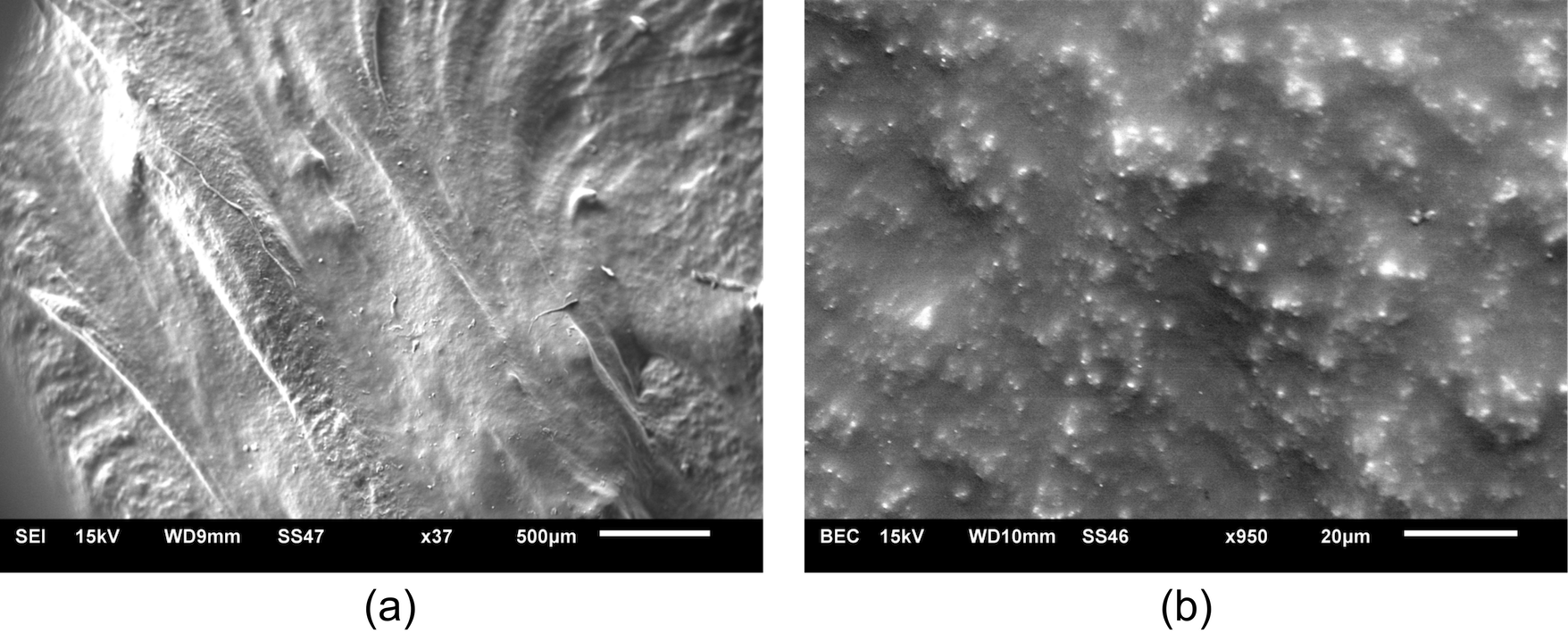}
    \caption{SEM images of the fabricated AB/PDMS electrode with scale bars of (a) 500 $\mu$m and (b) 20 $\mu$m.}
    \label{fig:SEM}
\end{figure}

Scanning electron microscope (SEM) images of the fabricated AB/PDMS electrodes were taken using a low vacuum SEM microscope (JEOL 6610LV). Fig. \ref{fig:SEM} (a) shows a secondary electron detection mode of imaging with a scale bar of 500 $\mu$m, showing the surface of the fabricated electrodes (after wearing a few times). Fig. \ref{fig:SEM} (b) shows a backscattered electron detection mode of imaging with a scale bar of 20 $\mu$m, showing the AB particles. 
 
\begin{figure}[!ht]
    \centering
    \includegraphics[width=.95\linewidth]{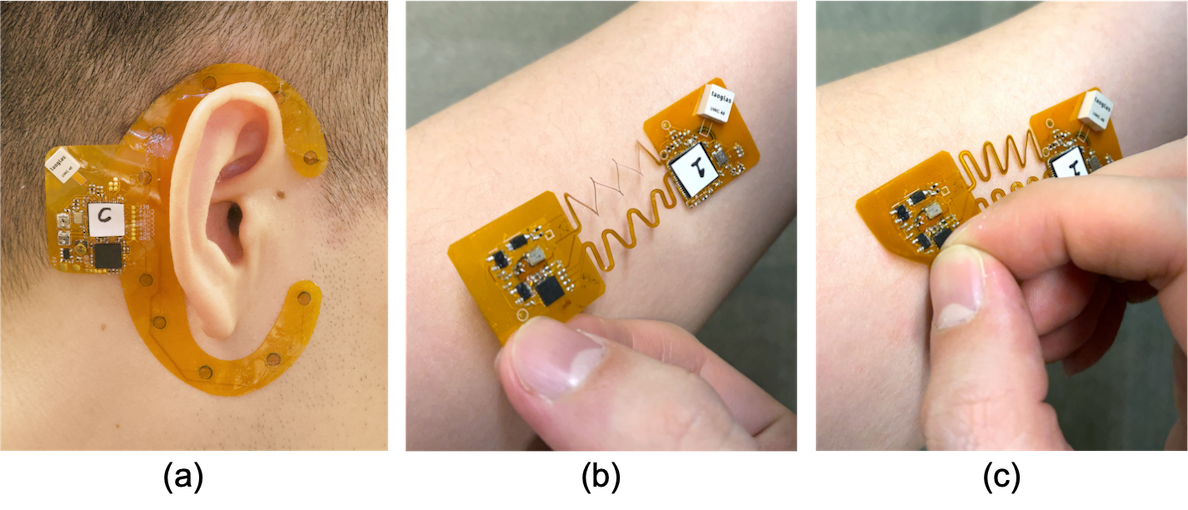}
    \caption{Photos of (a) the C-shape central node, and the patch-type sensor node during (b) stretching, and (c) peeling off.}
    \label{fig:device}
\end{figure}

Fig. \ref{fig:device} shows the devices on a user. Thanks to the flexible substrate and the gel-free electrodes, the devices can be reliably and comfortably worn for a long time. The non-conductive areas have a strong adhesion property, which is essential for forming a stable sensor-skin contact. The flexible sensors can also be easily peeled off without leaving residues on the user's skin. The serpentine shape structure allows 10-20\% deformation, which is sufficient for the contraction and relaxation processes of the target muscle groups.

\subsection{Testing of the Electrodes}

\begin{figure}[!ht]
    \centering
    \includegraphics[width=.8\linewidth]{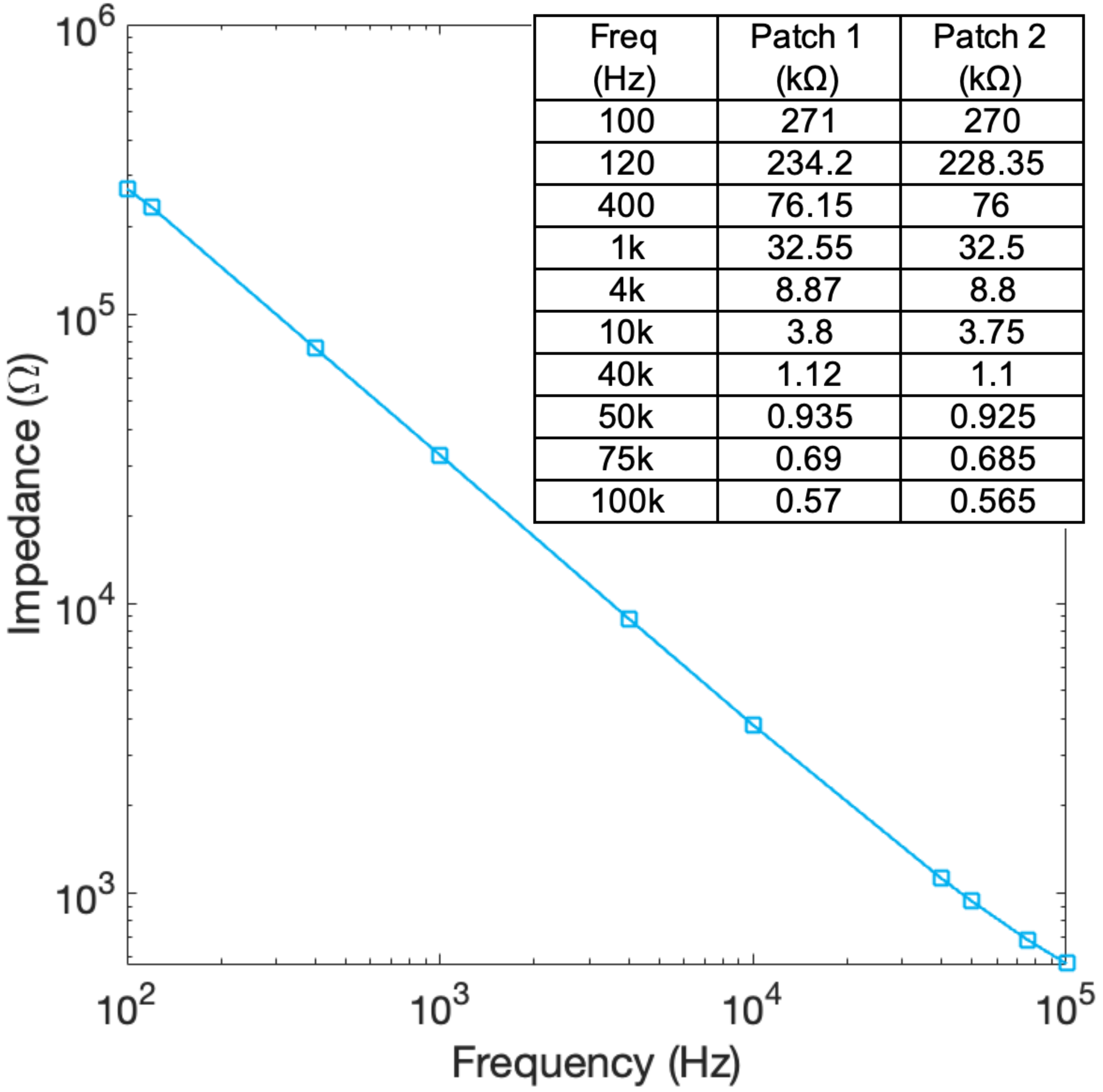}
    \caption{Measured impedance of the fabricated electrodes with AB (5~wt\%) in PDMS. The table shows the measurement from two patches.}
    \label{fig:exp_elec}
\end{figure}

{\color{black}The thickness of the electrode is about 1.8mm +/- 0.2mm. The electrode area is slighter thicker (approximately 0.2mm) than the adhesive area. These are experimentally selected parameters to ensure tight electrode contact while without affecting the stability of the sensor-skin contact.} The conductive impedance of the electrodes was measured on skin. The measurement results of two EMG patches are shown in Fig. \ref{fig:exp_elec}. The impedance shows a capacitive characteristic within the frequency range of interest. The impedance at 1 kHz is 32.5 k$\Omega$. 

\begin{figure}[!ht]
    \centering
    \includegraphics[width=1\linewidth]{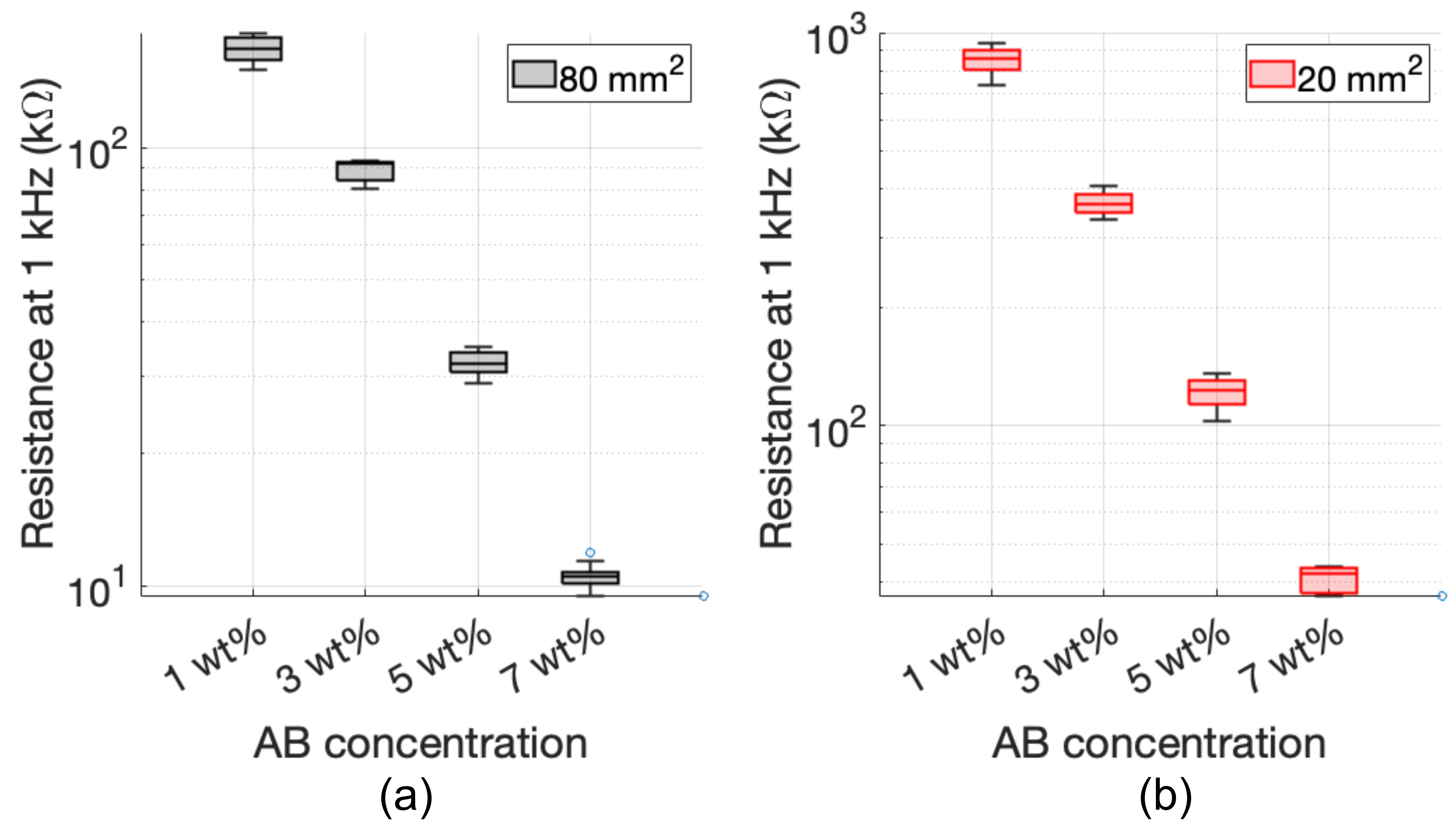}
    \caption{{\color{black}Measured resistance at 1 kHz with different AB concentrations for electrodes with surface areas of (a) 80 mm\textsuperscript{2} and (b) 20 mm\textsuperscript{2}, respectively.}}
    \label{fig:exp_res}
\end{figure}

{\color{black}The electrode resistances with varying AB concentrations from 1 wt\% to 7 wt\% were examined with electrode surface areas of 80 mm\textsuperscript{2} and 20 mm\textsuperscript{2}, respectively. The measurements were performed at 1 kHz, and the results of 10 electrodes were plotted in Fig.~\ref{fig:exp_res}. As anticipated, an increase in AB particle concentration leads to a decrease in resistance. Nonetheless, surpassing a concentration of 7~wt\% is impractical due to the challenges associated with stirring the mixture. The resistance for 7~wt\% electrodes is around 10 k$\Omega$ at 1 kHz with a surface area of 80 mm\textsuperscript{2}, which is comparable to gel electrodes that are commonly used in existing wearable devices.}

\subsection{Testing of the Flexible Sensors}

\begin{figure}[!ht]
    \centering
    \includegraphics[width=.95\linewidth]{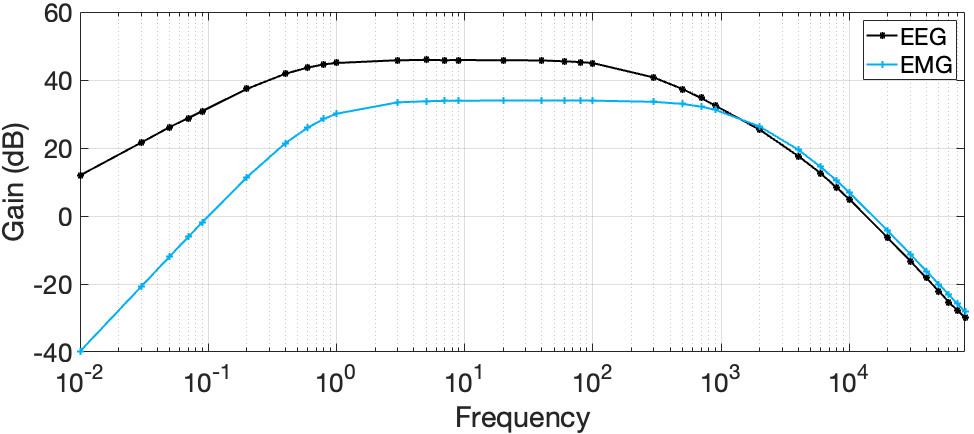}
    \caption{Measured frequency responses of the low noise amplifiers in the default configuration for EEG and EMG recording, respectively.}
    \label{fig:exp_freq}
\end{figure}

\begin{figure}[!ht]
    \centering
    \includegraphics[width=.95\linewidth]{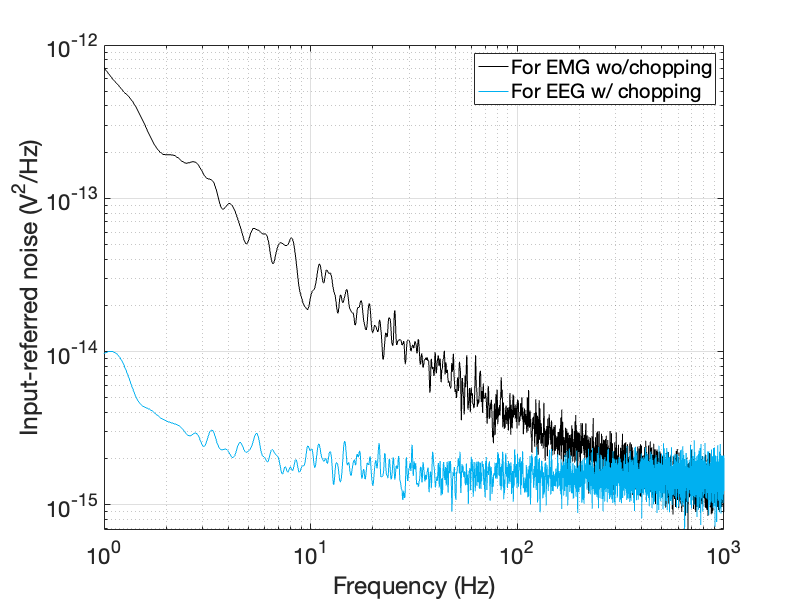}
    \caption{\color{black}The measured input-referred noise power spectrum of the low-noise instrumentation amplifier with and without chopping stabilization enabled for EEG and EMG recording, respectively.}
    \label{fig:exp_LNA}
\end{figure}

{\color{black}The measured frequency responses of the low-noise instrumentation amplifier with different configurations for EEG and EMG recording are shown in Fig.~\ref{fig:exp_freq}. The measured input-referred noise power spectrum of the low-noise instrumentation amplifier with and without chopping stabilization is shown in Fig.~\ref{fig:exp_LNA}.} 
The amplifiers at the central node were configured for EEG recording with a bandwidth of 0.5 to 200 Hz. The closed-loop gain was set to 200 and the measured mid-band gain was 199.2 (0.4\% gain error). The input-referred noise was 1.9~$\mu$V with chopper stabilization enabled. The calculated noise efficiency factor (NEF) is 4.72~\cite{harrison2003low}.
The amplifier in the patch-type sensor was configured for EMG recording. The bandwidth of the amplifier was 1 Hz to 1 kHz. The measured mid-band gain was 49.95 (ideal is 50) and the input-referred noise was 4.3~$\mu$V with chopper stabilization disabled. The calculated NEF is 3.12. The common-mode rejection ratio (CMRR) was measured to be -105 dB at 100 Hz. The input impedance was measured to be over 800 M$\Omega$ when the positive feedback loop was enabled.

\begin{figure}[!ht]
    \centering
    \includegraphics[width=1\linewidth]{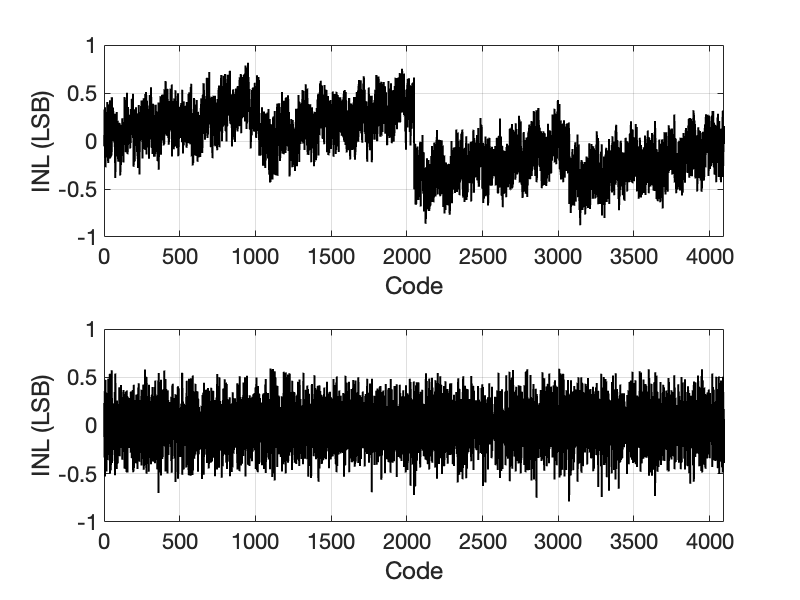}
    \caption{{\color{black}Measure INL and DNL of the developed 12-bit ADC.}}
    \label{fig:ADC_INL_DNL}
\end{figure}

The ADC was tested on the bench at a sampling rate of 100 kSps with an input tone at -0.5 dBFs near the Nyquist sampling rate. The measured spurious-free dynamic range (SFDR) was 78.1 dB and the signal to noise and distortion ratio (SNDR) was 67.4 dB. The calculated effective number of bits (ENOB) was 10.9-bit. The ENOB can be improved to 11.3-bit if the signal bandwidth is less than 10 kHz. The power figure of merit (FoM) of the ADC was calculated to be 61.3 fJ/conv-step. {\color{black}The integral nonlinearity (INL) and differential nonlinearity (DNL) were measured on the bench using a slow wave signal generated by a high precision DAC (AD5791, Analog Devices Inc.). Fig.~\ref{fig:ADC_INL_DNL} shows the measured INL and DNL versus the input codes. The measurement results show that the worst-case INL and DNL were less than 0.87 LSB and 0.79 LSB, respectively. }

\begin{figure}[!ht]
    \centering
    \includegraphics[width=.9\linewidth]{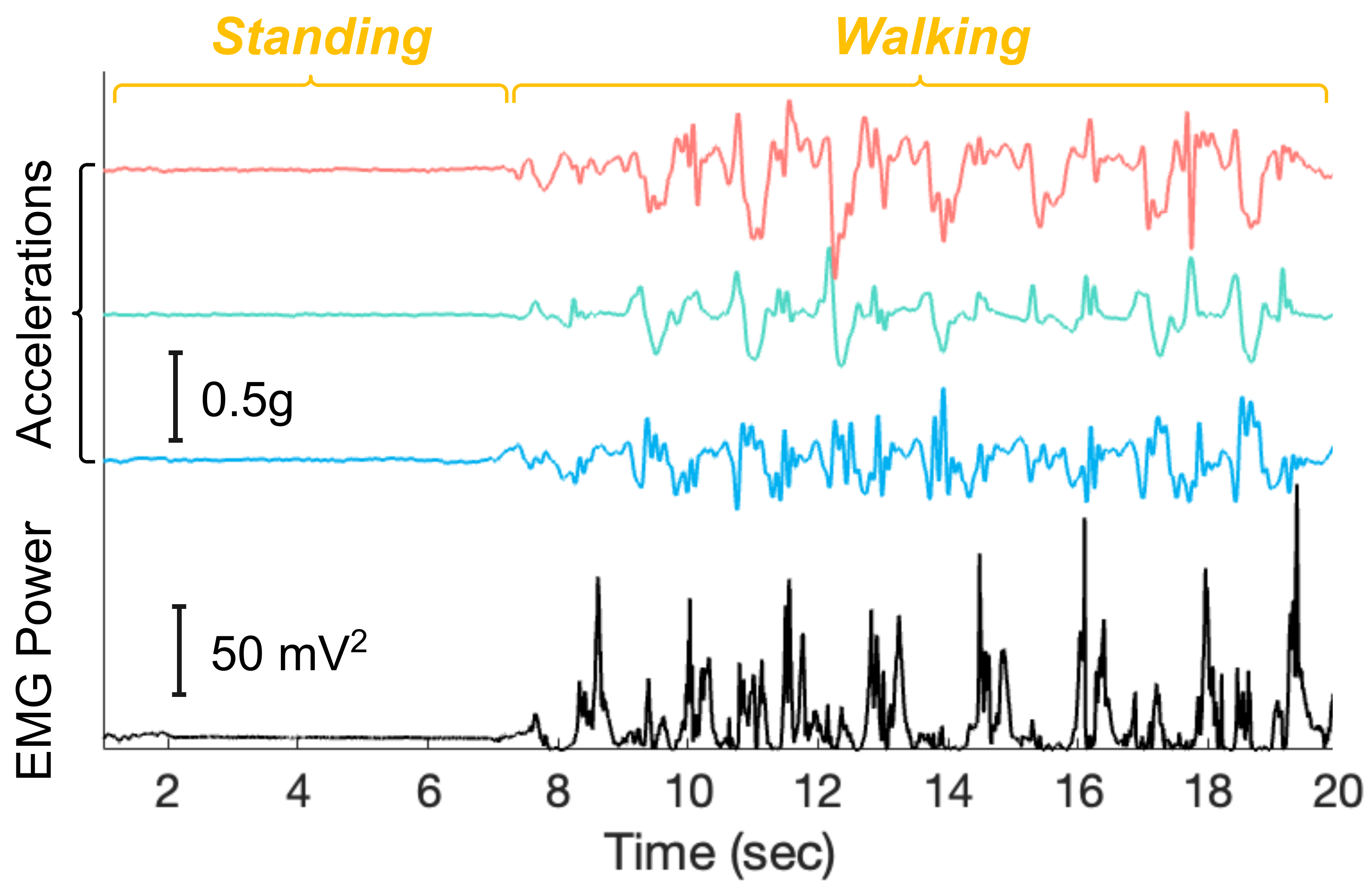}
    \caption{Measured EMG and 3-axis ACC signals when the user is standing and walking.}
    \label{fig:exp_EMG_acc}
\end{figure}

The neural front-end of the patch-type sensors was tested in users while standing and walking. Fig. \ref{fig:exp_EMG_acc} shows the measured EMG and ACC signals from the custom IC. Signal characteristics during standing and walking are visible in the recording. 

\begin{figure}[!ht]
    \centering
    \includegraphics[width=.8\linewidth]{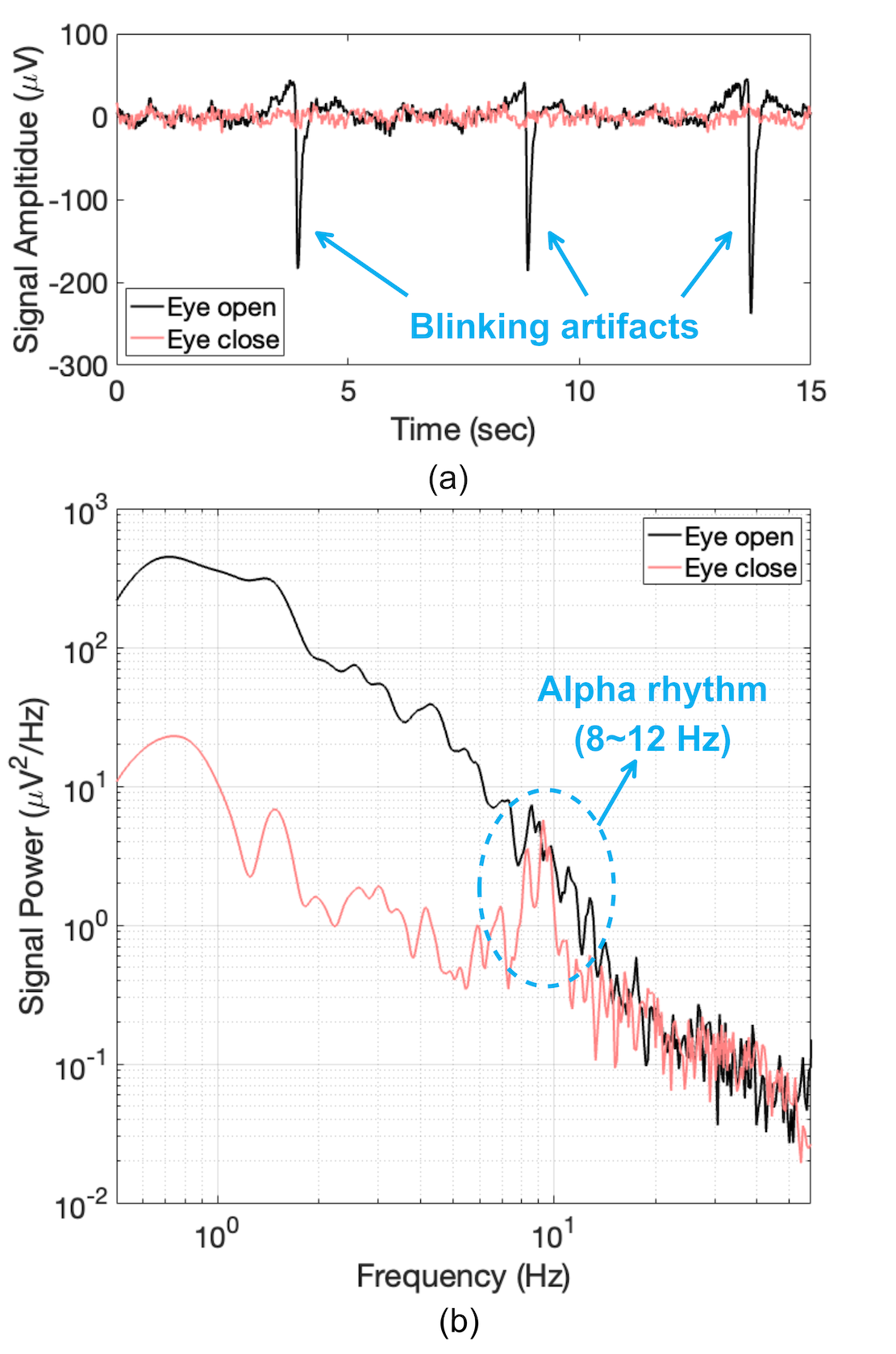}
    \caption{Measured EEG signal from the C-shape central node. (a) Measured time-domain EEG signals with eyes open and closed. Blinking artifacts are visible when the subject eyes are open. (b) Power spectrum of the EEG signals with eyes open and closed. The alpha rhythm (8-12 Hz) is visible when the subject eyes are closed.}
    \label{fig:exp_EEG}
\end{figure}

The EEG recording performance of the central sensor has also been validated on users. Fig. \ref{fig:exp_EEG} shows the measured EEG signal when the user kept the eyes open and closed. The blinking artifacts are visible in the recording, as shown in Fig. \ref{fig:exp_EEG}(a), which can be removed by post-processing. The power spectrum of the recording when the user kept the eyes closed shows an activation of the alpha rhythm, which is the brain oscillation around 8-12 Hz, as expected \cite{sigala2014role}. 
 
\begin{figure}[!ht]
    \centering
    \includegraphics[width=.9\linewidth]{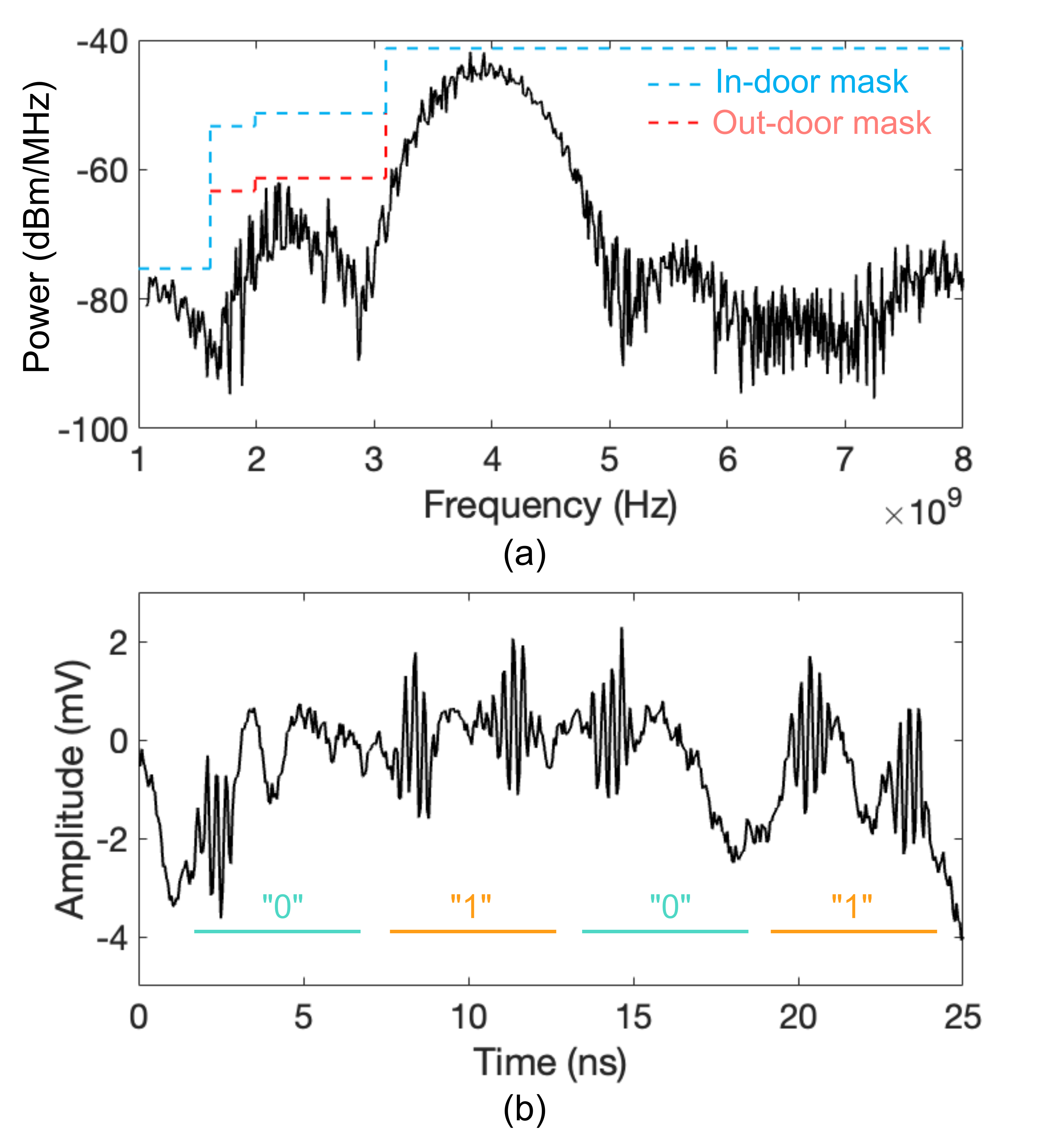}
    \caption{(a) The measured output spectrum of the UWB TX (RBW = 1MHz). (b) The measured time-domain signal at the UWB RX end.}
    \label{fig:exp_UWB}
\end{figure}

The measured power spectrum of the UWB TX is shown in Fig. \ref{fig:exp_UWB}(a). The TX output power is in compliance with the Federal Communications Commission (FCC) regulation for both outdoor and indoor operations, as marked in the figure \cite{breed2005summary}. The TX output power can be reduced for power savings. Fig. \ref{fig:exp_UWB}(b) shows the time domain signal at the RX end measured using a high sampling rate oscilloscope. The bit error rate (BER) was tested with the RX decoder. With a five-pulse averaging scheme \cite{soltani202221}, the measured BER was better than $10^{-6}$ at a distance of 1.5 m with a data rate of 40 Mbps. This data rate is much higher than the required throughput of the application and the transceivers can be put in sleep mode for most of the time when not transmitting. For hardware simplicity and power saving, the multiple patch-type sensors transfer data using a time division scheme where all sensor nodes are synchronized initially. An identification header was added to each sensor node for discrimination. The measured power consumption for transmission at 40~Mbps is 3.4~pJ/bit for the TX and 110.7 pJ/bit for the RX.

The power consumption of the central sensor was measured to be 17 mA on average during operation, including 11.8 mA taken by the MCU for inference of the DL model. The power of the custom IC was dominated by the UWB RX. The central sensor is powered by a lithium battery. The power consumption of the patch-type sensor was 0.74~mA during continuous operation. This power can be provided by wireless inductive coupling with the transmitter coil placed under the feet of the users, for example, in shoes where batteries can be placed. {\color{black} The measured inductance of the fabricated F-PCB coil is 55 $\mu$H, and the quality factor is 2.3. A resonant capacitor in parallel with the inductive coil was experimentally selected to maximize the resonance at the operational frequency of 150~kHz. A 60~$\mu$H coil with a diameter of 38 mm was used as the transmitter coil. Experimental results show that if two lithium batteries (each with a typical voltage of 3.7~V) connected in series are used as the power source, the patch-type sensor can be wirelessly powered at a distance of 17.5~cm. If three lithium batteries are used in series, the wireless powering distance can be extended to 22.1~cm. Although wireless power transfer has limited power efficiency, it offers an alternative method to supply power to patch-type sensor nodes and allows for the reduction of sensor weight and thickness. This, in turn, enhances the comfort level for prolonged use. }

\subsection{Testing of the DL Model}

The developed DL model was evaluated using a leave-one-out (LOO) cross-testing strategy \cite{liu2021edge}. The model was tested on the data from each of the subjects on a rotating basis, while the testing data is not seen during the training and validation. We benchmarked the inference performance of the DL model using a single signal modality, including EMG, EEG, ACC, and using a combination of multi-modal sensory inputs. The fully connected layers are re-trained for each case.

\begin{figure}[!ht]
    \centering
    \includegraphics[width=.9\linewidth]{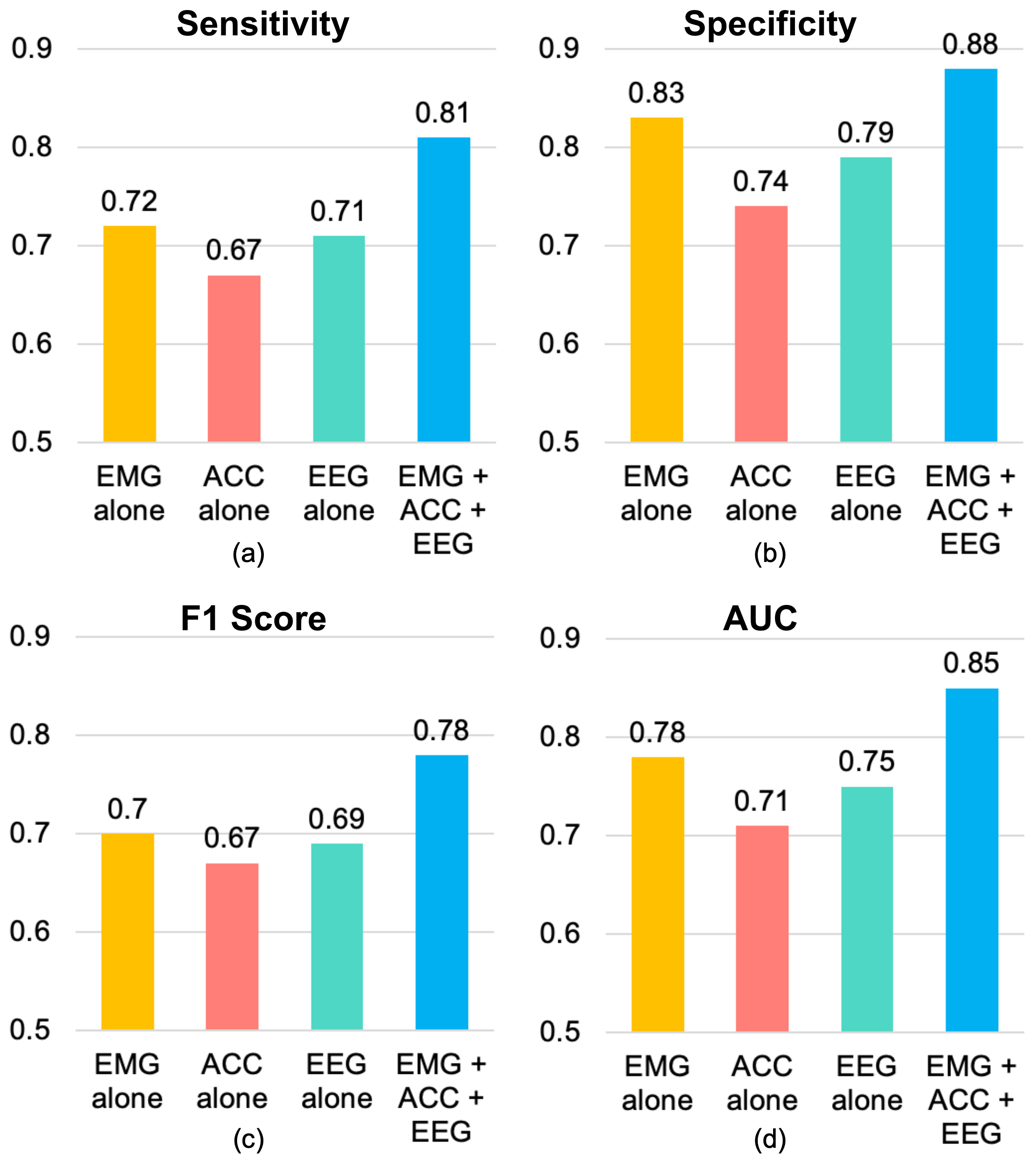}
    \caption{The average performance metrics of the developed DL model using single and multiple sensing modalities with 8-bit quantization, including (a) sensitivity, (b) specificity, (c) F1 score, and (d) AUC. }
    \label{fig:dl_performance}
\end{figure}

Sensitivity, specificity, F1 score, and area under the receiver operating characteristic curve (AUC) are used to evaluate performance \cite{huang2005using,mitchell1997machine}. An 8-bit quantization was applied to reduce the model size and computational costs by avoiding floating point computing \cite{liu2021edge}. Performance degradation due to quantization is less than 1\% on average. The key performance of the model after 8-bit quantization is summarized in Fig. \ref{fig:dl_performance}. The testing results suggest that the model performance using multi-modal sensory inputs surpasses those using a single signal modality: the final F1 score is 0.78, which is 8\% higher than the best score using a single modality; the final AUC is 0.85, which is 7\% higher than the best score using a single modality. The multi-modal performance of this work is comparable to the FoG detection results in state-of-the-art works that were not implemented on hardware \cite{zhang2022multimodal,bajpai2022multi,guo2022high}. The measured total inference time using multi-modal sensory input is less than 2.3~s, which is less than the 3~s input window. 

{\color{black}Table 1 compares the performance of our work with state-of-the-art approaches. It's worth noting that all other methods listed in the table were developed without any limitations on computational resources. Despite being constrained by the hardware resources of a low-power microcontroller, our model achieved levels of performance comparable and superior to the best-reported results, especially in overall accuracy.}

\begin{table*}[!ht]
\renewcommand{\arraystretch}{1.1} 
\centering
\caption{{\color{black}Comparison with State-of-the-art Works}}\label{table:comparison}
\begin{tabular}{|c|c|c|c|c|c|c|}
\hline
Reference                  & Sensor      & Methods & Sensitivity & Specificity & F1-score & Accuracy \\ \hline
2013 \cite{handojoseno2013using} & EEG & Multilayer perception (MLP) & 0.82  & 0.6  & - & 0.71 \\ \hline
2015 \cite{mazilu2015prediction} & ECG + SC & Multivariate Gaussian distribution (MGD) & - & - & - & - \\ \hline
2017 \cite{naghavi2019prediction} & IMU         & Linear Discriminant Analysis (LDA)     & 0.83        & 0.67        & -       & 0.71 \\ \hline
2018 \cite{sama2018determining} & IMU & Support vector machine (SVM) & 0.96 & 0.71 & - & 0.75 \\ \hline
2022 \cite{guo2022high} & IMU & SVM & 0.75 & - & 0.74 & 0.83 \\ \hline
This work                  & IMU+EMG+EEG & Depth-wise CNN     & 0.81        & 0.88        & 0.78   & 0.85  \\ \hline
\end{tabular}
\end{table*}

\begin{table}[!ht]
\renewcommand{\arraystretch}{1.1} 
\caption{Measured Performance of the Flexible Sensors}\label{table_specs}
\centering
\begin{tabular}{|l|l|l|}
\hline
\multicolumn{1}{|c|}{\textbf{Module}}                                                                 & \multicolumn{1}{c|}{\textbf{Items}} & \multicolumn{1}{c|}{\textbf{Performance}}                                                                                     \\ \hline
\multicolumn{1}{|c|}{\multirow{5}{*}{\begin{tabular}[c]{@{}c@{}}Low-noise\\ Amplifiers\end{tabular}}} & Noise                               & \begin{tabular}[c]{@{}l@{}}1.9 $\mu$V for EEG\\ (0.5-200 Hz w/ chopping)\\ 4.3 $\mu$V for EMG\\ (1-1 kHz wo/ chopping)\end{tabular} \\ \cline{2-3} 
\multicolumn{1}{|c|}{}                                                                                & NEF                                 & \begin{tabular}[c]{@{}l@{}}4.72 for EEG\\ 3.12 for EMG\end{tabular}                                                           \\ \cline{2-3} 
\multicolumn{1}{|c|}{}                                                                                & Mid-band gain                       & \begin{tabular}[c]{@{}l@{}}199.2 for EEG\\ 49.95 for EMG\end{tabular}                                                         \\ \cline{2-3} 
\multicolumn{1}{|c|}{}                                                                                & CMRR                                & \textgreater 105 dB                                                                                                           \\ \cline{2-3} 
\multicolumn{1}{|c|}{}                                                                                & Input Impedance                     & \textgreater 800 M$\Omega$                                                                                                         \\ \hline
\multirow{4}{*}{ADC}                                                                                  & Sampling rate                       & up to 100 kSps                                                                                                                \\ \cline{2-3} 
                                                                                                      & ADC power                           & 61.3 fJ/conv-step                                                                                                             \\ \cline{2-3} 
                                                                                                      & ADC ENOB                            & 10.9-bit                                                                                                                      \\ \cline{2-3} 
                                                                                                      & SNDR/SFDR                           & 67.4 dB/78.1 dB                                                                                                               \\ \hline
\multirow{4}{*}{UWB}                                                                                  & TX output power                     & up to -6 dBm                                                                                                                  \\ \cline{2-3} 
                                                                                                      & Data rate                           & 40 Mbps                                                                                                                       \\ \cline{2-3} 
                                                                                                      & BER                                 & \textless 10\textasciicircum{}-6 at 1.5 m                                                                                               \\ \cline{2-3} 
                                                                                                      & UWB power                           & \begin{tabular}[c]{@{}l@{}}3.4~pJ/bit for TX\\ 110.7~pJ/bit for RX\end{tabular}                                               \\ \hline
\multirow{4}{*}{DL model}                                                                             & Sensitivity                         & 0.81 (0.793, 0.825)*                                                                                                                       \\ \cline{2-3} 
                                                                                                      & Specificity                         & 0.88 (0.856, 0.91)*                                                                                                                       \\ \cline{2-3} 
                                                                                                      & F1 score                            & 0.78 (0.739, 0.811)*                                                                                                                       \\ \cline{2-3} 
                                                                                                      & AUC                                 & 0.85 (0.813, 0.887)*                                                                                                                       \\ \hline
\multirow{4}{*}{System}                                                                               & Technology                          & \begin{tabular}[c]{@{}l@{}}Polyimide Flex-PCB\\ 180 nm CMOS\end{tabular}                                                      \\ \cline{2-3} 
                                                                                                      & Supply voltages                     & 1.8 V/1.2 V                                                                                                                   \\ \cline{2-3} 
                                                                                                      & Supply current                      & \begin{tabular}[c]{@{}l@{}}17 mA for central node\\ 740 uA for patch-type nodes\end{tabular}                                  \\ \cline{2-3} 
                                                                                                      & Power plan                          & \begin{tabular}[c]{@{}l@{}}Li-battery for the central node\\ WPT for patch-type nodes\end{tabular}                                \\ \hline
\end{tabular}
\begin{flushleft}\hspace{5mm}* 95\% confidence interval (CI).\end{flushleft}
\end{table}

\section{Discussion and Future Work}

In this work, we developed a wireless flexible sensor network for real-time FoG detection and alert. The sensor network consists of two types of wearable sensors with gel-free electrodes. Custom IC and general-purpose electronic components were used to meet the system specifications for this application. The key performance of the system is summarized in Table \ref{table_specs}. The developed sensors are proven to be comfortable to wear over the long term and provide reliable signal acquisition and processing capability. 

The size of the DL model is limited by the memory and computational resources of the selected MCU module. These limitations can be overcome by validating the model using a more powerful MCU module, such as the HX6537-A (Himax Technologies). Since this design is a proof-of-concept and the DL model may need to be adapted when conducting the experiments on PD patients, we have not tried more complicated models on more powerful MCU modules yet. Eventually, the DL model can be implemented in custom ICs to achieve optimal performance and power consumption. 

{\color{black}The overall project is divided into two phases. In phase one, we developed a wearable device capable of multi-modal signal acquisition and real-time ML, and demonstrated its ability to provide reliable, high-quality signals and comfortable wear over long periods. The results of this phase are reported in this paper, which does not include clinical trials. In phase two, the devices will be tested on patients at the Toronto Rehabilitation Institute (TRI), Toronto, ON, Canada. All procedures will obtain approvals from the research ethics board (REB) at the University Health Network (UHN), Canada, and written informed consent will be obtained from all participants. Experiments will be conducted in the off-medication state of patients. The second phase has two major objectives. The first objective aims to collect multi-modal data from PD patients, including both male and female patients. EEG recording will be performed with cap EEG headset and ear EEG device simultaneously for bench marking. The patients will be video recorded using motion and in-depth cameras to assist FoG event labeling. The labeled data will be used to train the ML model for real-time inference and auditory feedback generation. The second objective aims to statistically analyze the effectiveness of the real-time auditory feedback and compare it with control groups. This work provides a strong foundation for obtaining approval and continuing our research in the planned subsequent phases. 
}

\section{Conclusion}

This paper presents a first-of-its-kind wireless sensor network design for FoG detection and alerting. The developed system has three key advantages. First, the system is fully self-contained and the operation is autonomous, making it convenient for the users. Second, the sensors are flexible and comfortable to wear over the long term, and the electrodes are easy to fabricate at a low cost. Third, a novel DL model and custom IC design techniques are proposed and implemented to achieve the application-specific performance and ultra-low power consumption. The system can potentially be used for clinical studies for PD patients. The design methods can be used for a broad range of applications for the detection of neurological disorders and providing closed-loop treatment. 

\bibliographystyle{IEEEtran}
\bibliography{ref}

\ifCLASSOPTIONcaptionsoff
  \newpage
\fi

\begin{IEEEbiography}[{\includegraphics[width=1in,height=1.25in,clip,keepaspectratio]{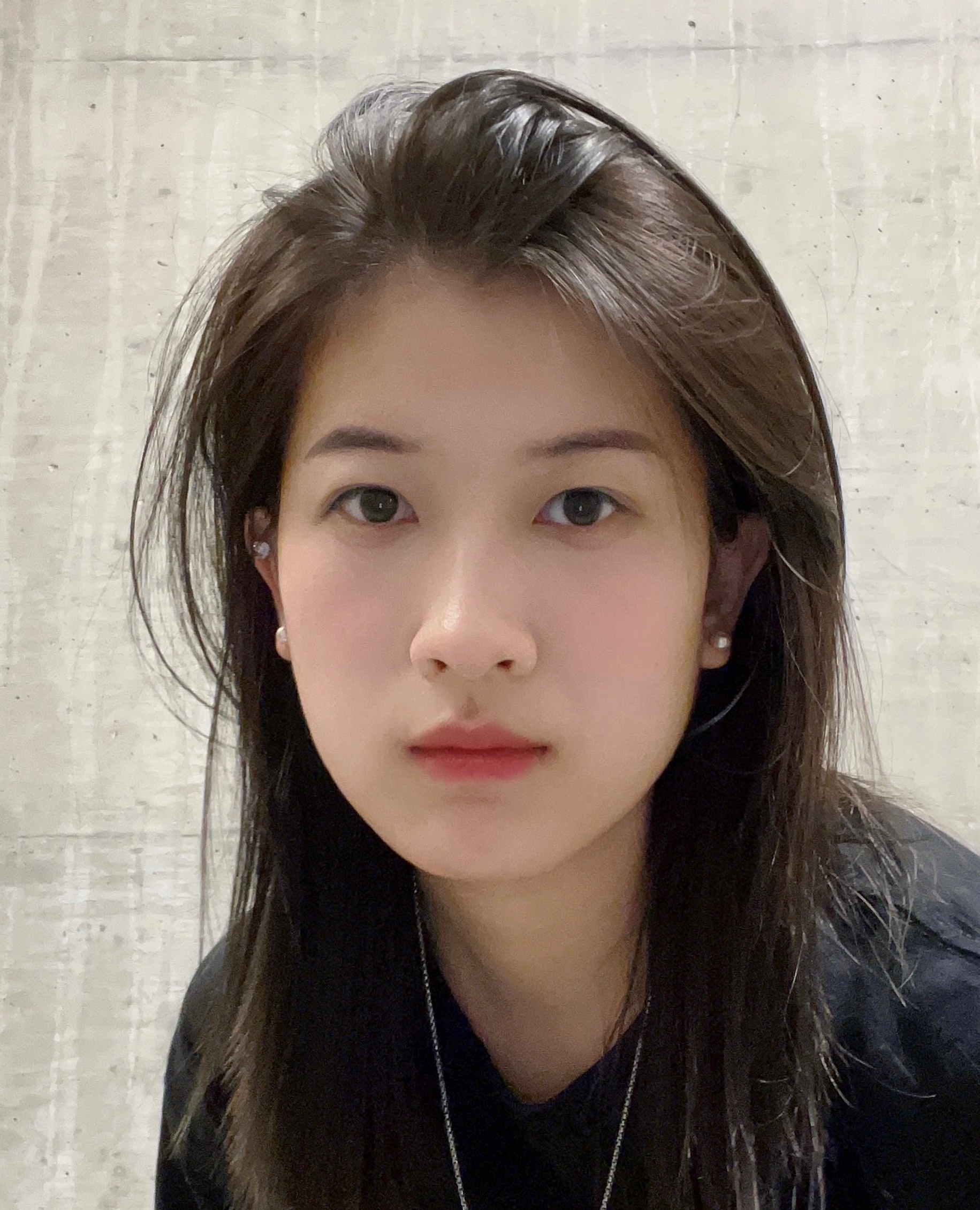}}]{Yuhan Hou} (Student Member, IEEE) is currently pursuing her master’s degree in Electrical and Computer Engineering at the University of Toronto, Toronto, ON, Canada. She received her Bachelor of Science degree in Electrical Engineering at the University of Toronto in 2023. Her research interests include miniatured electronics design, medical sensors, and IC design for biomedical use. She completed a 16-months internship at Alphawave Semi from May 2021 to August 2022 as a hardware validation engineer. She developed a webAPI and a GUI for data post processing and implemented multiple validation tests for silicon IPs.
\end{IEEEbiography}

\begin{IEEEbiography}[{\includegraphics[width=1in,height=1.25in,clip,keepaspectratio]{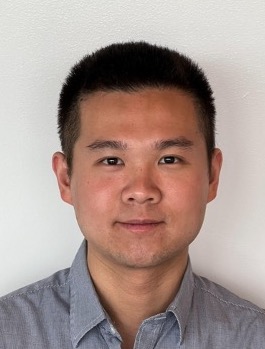}}]{Xing (Jack) Ji} received his Master of Engineering degree from the University of Toronto in 2022. He previously attended the University of Waterloo where he obtained his bachelor's degree in applied science. Since 2011, he has worked in the field of embedded software for numerous industry leading companies including Apple Inc., Qualcomm Inc., and Meta Platforms Inc., where he has led engineering teams to bring hardware projects from the discovery phase all the way to mass production. He is currently working on embedded software for the next generation of Oculus Virtual Reality devices. 
\end{IEEEbiography}

\begin{IEEEbiography}[{\includegraphics[width=1in,height=1.25in,clip,keepaspectratio]{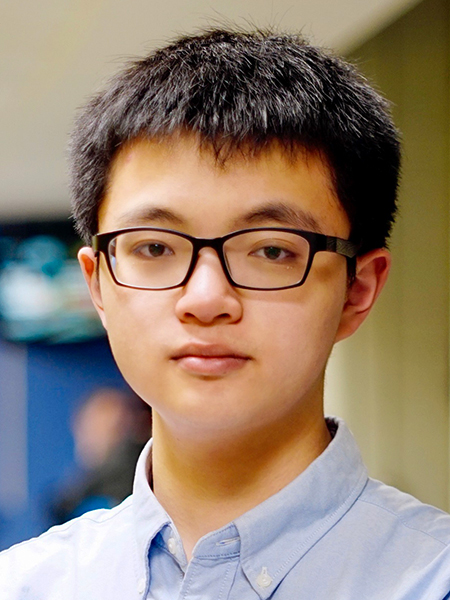}}]{Yi Zhu} (Member, IEEE) received the M.Eng. degree with an emphasis in Biomedical Engineering from The Edward S. Rogers Sr. Department of Electrical \& Computer Engineering at the University of Toronto, Toronto, ON, Canada, in 2022. Currently, he works as an Electrical Developer in the Surgical R\&D Hardware Department at Synaptive Medical Inc., a Toronto-based company focused on developing medical equipment and applications for neurosurgical procedures. Prior to joining Synaptive Medical Inc. in 2022, he was a graduate student in the X-Lab research team at the University of Toronto and a trainee at the KITE Research Institute, University Health Network. His research interests include analog and mixed-signal circuit designs for brain–machine interface and the next-generation robotic exoscope in Neurosurgery.
\end{IEEEbiography}

\begin{IEEEbiography}[{\includegraphics[width=1in,height=1.25in,clip,keepaspectratio]{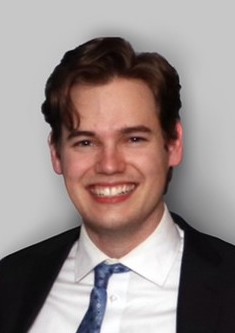}}]{Thomas Dell} was born in Oakville, Ontario, Canada in 1999.  He is currently studying to complete his BASc in Engineering Science with a major in Biomedical Systems Engineering at the University of Toronto in Toronto, Ontario, Canada.  Thomas completed his undergraduate thesis in 2022, which aimed to leverage Bluetooth Low Energy communication for use in a neural interfacing system.
\end{IEEEbiography}

\begin{IEEEbiography}[{\includegraphics[width=1in,height=1.25in,clip,keepaspectratio]{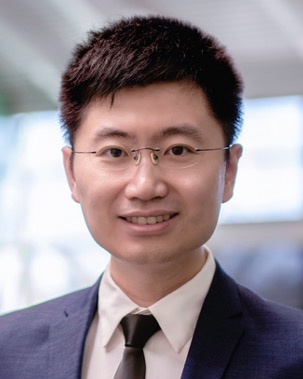}}]{Xilin Liu} (Senior Member, IEEE) obtained his Ph.D. degree from the University of Pennsylvania, Philadelphia, PA, USA, in 2017. 
He is currently an Assistant Professor in the Edward S. Rogers Sr. Department of Electrical and Computer Engineering (ECE) at the University of Toronto, Toronto, ON, Canada. He is also an affiliated scientist at the University Health Network (UHN), Toronto, ON, Canada. His research interests include analog and mixed-signal IC design for emerging applications in healthcare and communication. Before joining the University of Toronto in 2021, he held industrial positions at Qualcomm Inc., San Diego, CA, USA, where he conducted research and development of high-performance mixed-signal circuits for cellular communication. He led and contributed to the IPs that have been integrated into products in high-volume production. He was a visiting scholar at Princeton University, Princeton, NJ, USA, in 2014.

Dr. Liu received the Best Student Paper Award and the Best Track Award at the 2017 International Symposium on Circuits and Systems (ISCAS), the Best Paper Award (1st place) at the 2015 IEEE Biomedical Circuits and Systems Conference (BioCAS), the Best Track Award at the 2014 ISCAS, the student research preview (SRP) award at the 2014 IEEE International Solid-State Circuits Conference (ISSCC). His team was a finalist in the 2018 BCI Award and the 2022 BioCAS Grand Challenge. He also received the IEEE Solid-State Circuits Society (SSCS) Pre-doctoral Achievement Award at the 2016 ISSCC.
\end{IEEEbiography}

\end{document}